\documentclass[aps,reprint,superscriptaddress]{revtex4-1}
\usepackage{siunitx}
\usepackage{graphicx}
\usepackage{mathtools,amsthm}
\usepackage{epstopdf}
\usepackage{multirow}
\usepackage[normalem]{ulem}

\makeatletter
\AtBeginDocument{\let\LS@rot\@undefined}
\makeatother
\begin{document}
\title{Spin- and angle-resolved photoemission studies of the electronic structure of Si(110)``$16\times2$'' surfaces}
\author{N. K. Lewis}
\affiliation{School of Physics and Astronomy and the Photon Science Institute, University of Manchester, Oxford Road, Manchester, M13 9PL, UK}
\affiliation{The Cockcroft Institute, Daresbury Laboratory, Sci-Tech Daresbury, Warrington, WA4 4AD, UK}
\author{Y. Lassailly}
\affiliation{Physique de la Mati{\`e}re Condens{\'e}e, CNRS-Ecole Polytechnqiue, 91128 Palaiseau C{\'e}dex, France}
\author{L. Martinelli}
\affiliation{Physique de la Mati{\`e}re Condens{\'e}e, CNRS-Ecole Polytechnqiue, 91128 Palaiseau C{\'e}dex, France}
\author{I. Vobornik}
\affiliation{Istituto Officina dei Materiali (IOM), CNR, AREA Science Park Basovizza, I-34149 Trieste, Italy}
\author{J. Fujii}
\affiliation{Istituto Officina dei Materiali (IOM), CNR, AREA Science Park Basovizza, I-34149 Trieste, Italy}
\author{C. Bigi}
\affiliation{Istituto Officina dei Materiali (IOM), CNR, AREA Science Park Basovizza, I-34149 Trieste, Italy}
\affiliation{Department of Physics, Universit\`{a} degli Studi di Milano, I-20133 Milano, Italy}
\author{E. Brunkow}
\affiliation{Jorgensen Hall, University of Nebraska, Lincoln, Nebraska 68588-0299, USA}
\author{N. B. Clayburn}
\affiliation{Jorgensen Hall, University of Nebraska, Lincoln, Nebraska 68588-0299, USA}
\author{T. J. Gay}
\affiliation{Jorgensen Hall, University of Nebraska, Lincoln, Nebraska 68588-0299, USA}
\author{W. R. Flavell}
\affiliation{School of Physics and Astronomy and the Photon Science Institute, University of Manchester, Oxford Road, Manchester, M13 9PL, UK}
\author{E. A. Seddon}
\affiliation{School of Physics and Astronomy and the Photon Science Institute, University of Manchester, Oxford Road, Manchester, M13 9PL, UK}
\affiliation{The Cockcroft Institute, Daresbury Laboratory, Sci-Tech Daresbury, Warrington, WA4 4AD, UK}
\date{\today}
\begin{abstract}
The electronic structure of Si(110)``$16\times2$'' double-domain, single-domain and $1\times1$ surfaces have been investigated using spin- and angle-resolved photoemission at sample temperatures of $77\,\,\si{K}$ and $300\,\,\si{K}$. Angle-resolved photoemission was conducted using horizontally- and vertically-polarised $60\,\,\si{eV}$ and $80\,\,\si{eV}$ photons. Band-dispersion maps revealed four surface states ($S_1$ to $S_4$) which were assigned to silicon dangling bonds on the basis of measured binding energies and photoemission intensity changes between horizontal and vertical light polarisations. Three surface states ($S_1$, $S_2$ and $S_4$), observed in the Si(110)``$16\times2$'' reconstruction, were assigned to Si adatoms and Si atoms present at the edges of the corrugated terrace structure. Only one of the four surface states, $S_3$, was observed in both the Si(110)``$16\times2$'' and $1\times1$ band maps and consequently attributed to the pervasive Si zigzag chains that are components of both the Si(110)``$16\times2$'' and $1\times1$ surfaces. A state in the bulk-band region was attributed to an in-plane bond. All data were consistent with the adatom-buckling model of the Si(110)``$16\times2$'' surface. Whilst room temperature measurements of $P_y$ and $P_z$ were statistically compatible with zero, $P_x$ measurements of the enantiomorphic A-type and B-type Si(110)``$16\times2$'' surfaces gave small average polarisations of around 1.5\% that were opposite in sign. Further measurements at $77\,\,\si{K}$ on A-type Si(110)``$16\times2$'' surface gave a smaller value of +0.3\%. An upper limit of $\sim1\%$ may thus be taken for the longitudinal polarisation. 

\end{abstract}
\maketitle
\section{Introduction}
Spintronics encapsulates the generation, manipulation and detection of electron spins for use in devices primarily related to digital (binary) signal processing \cite{Zutic,Ando,Joshi}. Of key utility in such devices are ``spin transitions''. Considerable effort has been focused on the use of magnetic materials for injection of spin-polarised electrons into semiconductors, in the field called ``magneto-spintronics'' \cite{Datta,Dash, LeBreton}. However, impedance mismatching between a magnetic metal and a semiconductor represents a major design problem \cite{Schmidt,Hammar}. Theory shows that the higher-resistance semiconductor significantly depolarises the spin current from the ferromagnet unless the current is initially completely spin polarised. Several approaches have been taken to overcome this problem, including injection of electrons into the conduction band \cite{Jenson}, introduction of a tunnel contact between the semiconductor and ferromagnet \cite{Rashba}, and replacing the magnetic metals with a Heusler alloy \cite{Kar,Ishikawa}. Another approach is to use a semiconductor to generate the spin-polarised current \cite{Chuang}. Incorporation of silicon into spintronic devices is particularly important for compatibility with current CMOS technology \cite{Jenson}. Hence silicon is widely used as a substrate; the weak spin-orbit interaction is advantageous because it leads to spin coherence lengths of up to $1\,\,\si{\micro m}$ \cite{Sverdlov}, which allows manipulation of the spin current.  

Following pioneering work by a number of groups, non-magnetic surfaces are now well-known to give rise to spin separation, good examples being those of heavy metals \cite{Miyamoto,Yaji} and topological insulators \cite{Hsieh}. In both these cases the spin-orbit interaction is a key driver for the effect, coupled with the lack of inversion symmetry at surfaces. An additional property that can give rise to electron spin-polarisation effects is chirality; the transport of spin-polarised electrons through both random- and ordered-arrays of chiral molecules has been investigated \cite{Dreiling,Naaman}. Experiments probing the scattering of a transmitted electron beam through an enantiomerically-pure chiral target vapour have shown that the sign of the transmission asymmetry inverts upon changing the target molecule handedness \cite{Mayer}. This inversion was elucidated by the earlier theoretical work of Farago \cite{Farago}. The ordered enantiomers R,R and S,S 2-diphenyl-1,2-diethanediol adsorbed onto in-plane-magnetised Co thin films gave results that showed electrons spin polarised in their initial state (before photoexcitation) cannot only be changed in magnitude but also in direction after passage through chiral layers of adsorbates \cite{Nino}. In addition, this study revealed that complications may occur due to differences in adsorption geometry between enantiomers. Of particular note with reference to adsorbates is the pioneering work of Naaman and coworkers on spin-filtering through double stranded DNA oligomers that has been shown to give polarisations of between 50 and 60\% \cite{RaySG,Gohler}. The potential importance of these findings for spintronics applications was clearly recognised, but the molecular adsorbate/semiconductor combination is sub-optimal for technological applications. Given this we decided to investigate the surface electronic structure of a chiral reconstruction of silicon using spin-resolved photoemission.

In order to inform our experimental photoemission studies, semi-relativistic model calculations were performed upon two-dimensional lattices with and without mirror symmetry \cite{Modelling}. These showed that for a non-magnetic two-dimensional lattice without mirror symmetry (\textit{i.e.} a chiral lattice), there is a non-zero component of the spin polarisation that is ordinarily zero for lattices with mirror symmetry and that this component inverts between enantiomorphs. The orientation of this inverting component is parallel to the initial-state crystal momentum of the electron and is thus known as the longitudinal component. 

To determine experimentally if a chiral surface results in spin-polarised electrons without the need for an adsorbed chiral layer, our experimental starting point was to undertake spin- and angle-resolved photoemission from the well-studied chiral Si(110)``$16\times2$'' surface \cite{Lewis}. Although the Si(110)``$16\times2$'' reconstruction has been investigated using photoemission on a number of occasions \cite{Cricenti2,Kim,Sakamoto}, this paper discusses the first experimental investigation of the electronic structure of the chiral Si(110)``$16\times2$'' reconstruction using spin-resolved photoemission. Furthermore, new angle-resolved photoemission results are reported in which the incident photon energy and polarisation were varied as well as the surface temperature and morphology. Previous low-resolution double-domain band-dispersion maps (binding energy, $E_B$ against $k_{||}$) have been obtained by Crincenti \textit{et al.} \cite{Cricenti1,Cricenti2}. To investigate differences in angle-resolved photoemission measurements between double and single domains, we report here, for the first time, high-resolution double-domain band-dispersion maps. Our angle-resolved photoemission results build upon the work of Sakamoto \textit{et al.} \cite{Sakamoto} and Kim \textit{et al.} \cite{Kim} using different light polarisations to investigate the surface states of a single domain. This provides new information about the bonding type that has not been previously reported. We also contribute to the debate on the Si(110)``$16\times2$'' atomic arrangement by showing that our angle-resolved photoemission results are consistent with the AB model.

\section{The Si(110)``$16\times2$'' reconstruction}
\begin{figure}
	\centering
	\includegraphics[width=\linewidth]{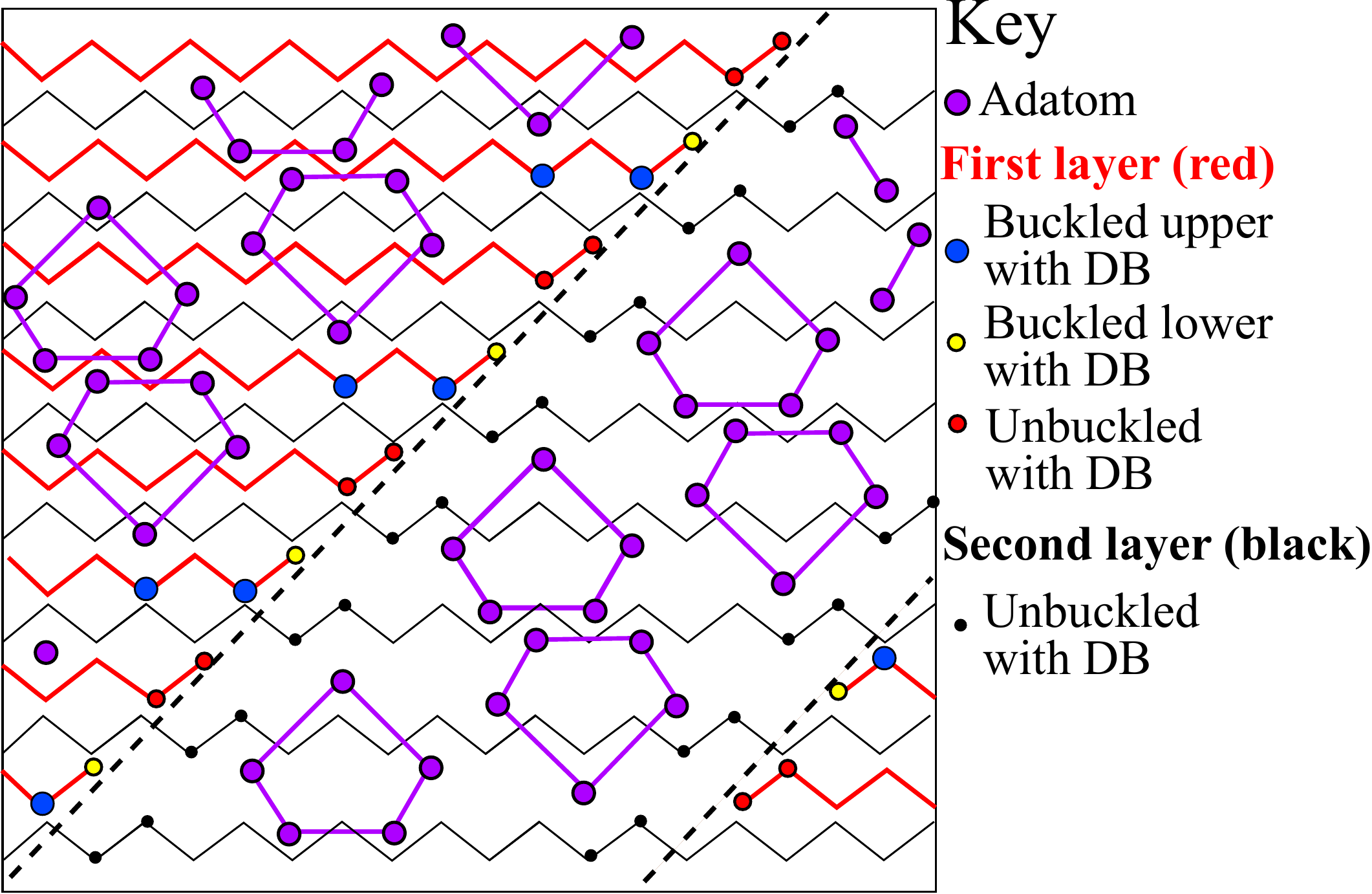}
	\caption{Schematic diagram of the adatom-buckling (AB) model that represents the Si(110)``$16\times2$'' reconstruction. The adatoms (shown by the purple circles) are positioned on top of the first (red, upper) and second (black, lower) layers and have dangling bonds (DBs). They are arranged into deformed pentagons which are indicated by the connecting lines. The first layer contains both unbuckled atoms and buckled-upper and buckled-lower atoms (shown by the large blue and the small red and yellow circles) all with DBs; all three atom types are located at the terrace edges. The second layer has dangling bonds that are located only on unbuckled atoms between the terrace edges and the adatoms \cite{Sakamoto}. The dashed lines correspond to the step edges and are parallel to either $[\bar{1}12]$ or $[1\bar{1}2]$ for a single-domain sample.}
	\label{Fig: AB model}
\end{figure}

On reconstruction the Si(110)``$16\times2$'' surface can exist either as a single domain consisting of only one enantiomorph (over several mm) or as a double domain with small areas of each enantiomorph \cite{An,Yamada,Lewis}. The reconstruction consists of a corrugated terrace structure where both upper and lower terraces have widths of $2.5\,\,\si{nm}$ and heights of $0.15\,\,\si{nm}$ \cite{Lewis}, the step edge of the corrugated terrace structure is parallel to either $[\bar{1}12]$ or $[1\bar{1}2]$ for a single-domain sample. On top of both the upper and lower terraces there are silicon atoms arranged into pairs of pentagons \cite{An}. In both single and double domains, the underlying (110) planes are formed of bonded silicon atoms that are described as zigzag chains \cite{An,Stekolnikov2,Harrison}. The exact atomic arrangement of the Si(110)``$16\times2$'' reconstruction is still under debate and several structural models have been proposed \cite{An,Setvin,Stekolnikov,Yamasaki}. Stekolnikov \textit{et al.} suggested the adatom-tetramer-interstitial model (ATI) \cite{Stekolnikov} to describe the Si(110)``$16\times2$'' reconstruction, which is no longer accepted because simulated scanning-tunnelling microscopy (STM) images from it are incompatible with experimentally obtained STM images \cite{Sakamoto}. Currently, the adatom-buckling (AB) model is the preferred structural picture because it has been shown to be consistent with both STM and Si $2p$ Auger spectroscopy measurements on a single-domain surface \cite{Kakiuchi}. A schematic diagram of the AB model is presented in Fig. \ref{Fig: AB model}. This model consists of adatoms, shown by the purple circles, that are arranged into an approximately pentagonal structure positioned on both the upper and lower terraces. Each adatom has a dangling bond, and there are three other types of atoms located at the step edges also with dangling bonds (DBs): unbuckled atoms (shown by red circles), buckled-upper atoms (shown by blue circles), and buckled-lower atoms (shown by yellow circles). The last set of atoms with DBs are those that reside on the lower terraces in between the step edge and the adatoms. In contrast, the Si(110)$1\times1$ surface consists of a single exposed layer of zigzag chains containing a single type of dangling bond \cite{Stekolnikov} which is described by the rotational-relaxation model \cite{Menon}.

\section{Experimental method}
\subsection{Sample reconstruction}
\begin{figure}
	\centering
	\includegraphics[width=0.85\linewidth]{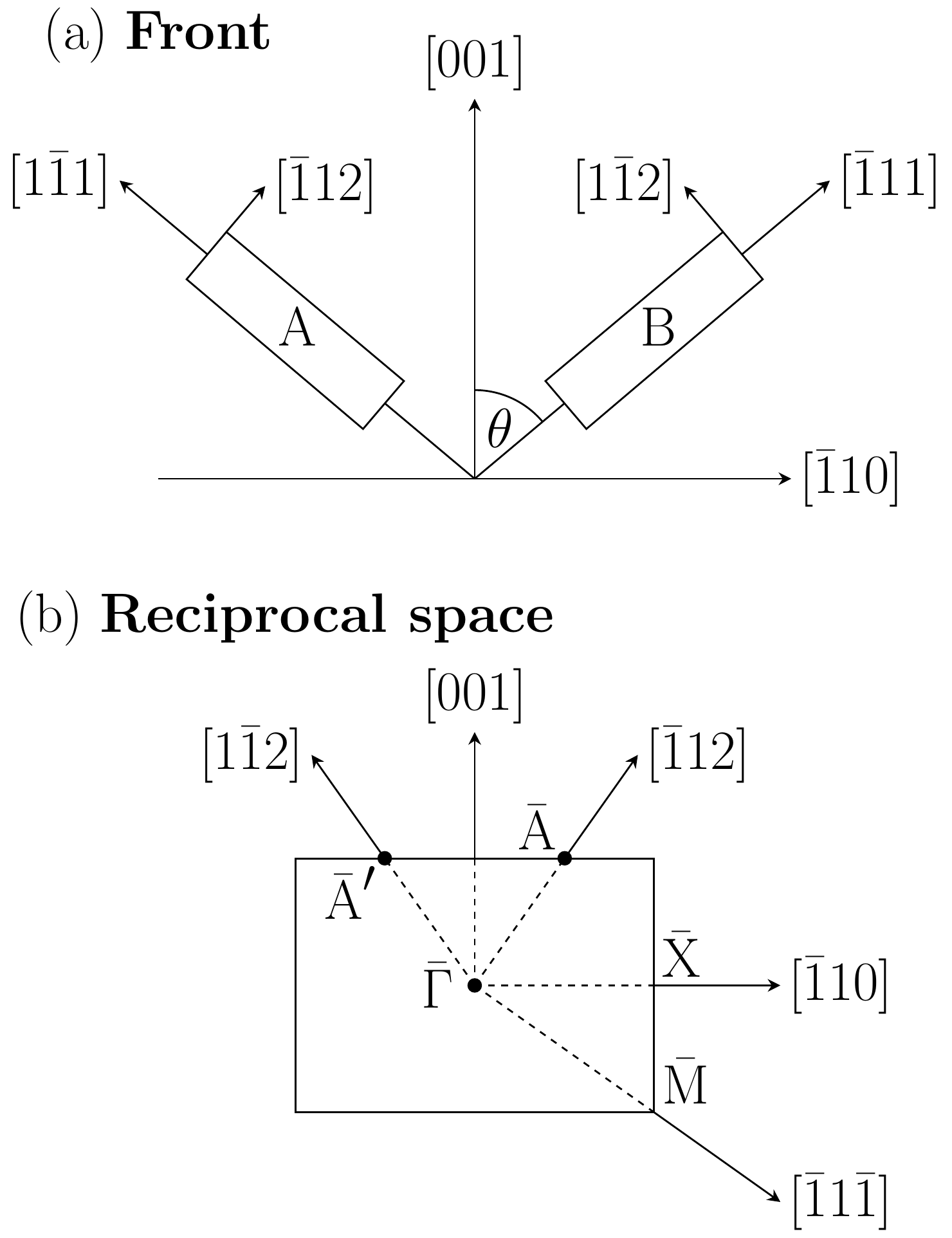}
	\caption{(a) Schematic diagram of the `front' faces \cite{Lewis} of the A and B Si(110) sample types used ($\theta=54.7\,\si{\degree}$). (b) Surface Brillouin zone for the Si(110)$1\times1$ surface showing the high symmetry points $\bar{\text{X}}$ and $\bar{\text{M}}$. The directions shown in (b) correspond to the real-space directions in (a).}
	\label{Fig1}
\end{figure}

Two types of $7\times2\,\,\si{mm}$ Si(110) samples were used, labelled as A or B where the short axis was parallel to the $[\bar{1}12]$ or $[1\bar{1}2]$ directions, respectively \cite{Lewis}. The silicon wafers were phosphorus doped giving a resistivity of $4-6\,\,\Omega\,\text{cm}$ and were supplied by PI-KEM Ltd. and SurfaceNet GmbH. Figure \ref{Fig1}(a) shows the relative orientation and `front' face, as defined in Ref. \cite{Lewis}, of the two Si(110) sample types. Only the `front' faces were used in this experiment; these were identified by generating the $1\times1$ surface and relating the observed LEED pattern with the direct-space lattice vectors.

The chiral Si(110)``$16\times2$'' reconstruction was generated by resistively heating the samples using a direct current parallel to the sample long axis in an UHV preparation chamber with a base pressure of $2\times10^{-10}\,\,\si{mbar}$. The samples were outgassed at $650\,\si{\degreeCelsius}$ for approximately 12 hours, then flashed several times to $1200\,\si{\degreeCelsius}$. After the final flash the samples were annealed for 30 seconds at $720\,\si{\degreeCelsius}$ and then cooled to room temperature by reducing the current $30\,\,\si{mA}$ every 30 seconds. LEED was used to determine the type of Si(110)``$16\times2$'' reconstruction generated. Si(110)$1\times1$ surfaces were produced by quenching the sample directly from $1200\,\si{\degreeCelsius}$ to room temperature.

The handedness of the samples was determined \textit{a posteriori} and before analysis by photoemission spectroscopy. This is because the front of each individual sample type (A and B) was polished such that the step-edge direction of the vicinal structure observed at high temperature \cite{Yamamoto} was parallel to the short axis of the sample. Upon successful generation of a single domain, the reconstruction handedness is known because the step orientation of the vicinal structure causes the corrugated terraces to be parallel to the short axis. Therefore, a single domain on the front face of A-type or B-type samples produces only an L or R domain, as defined by Yamada \textit{et al.} \cite{Yamada}, over mm dimensions.

\subsection{Spin- and angle-resolved photoemission}
\begin{figure}[!htbp]
	\centering
	\includegraphics[width=\linewidth]{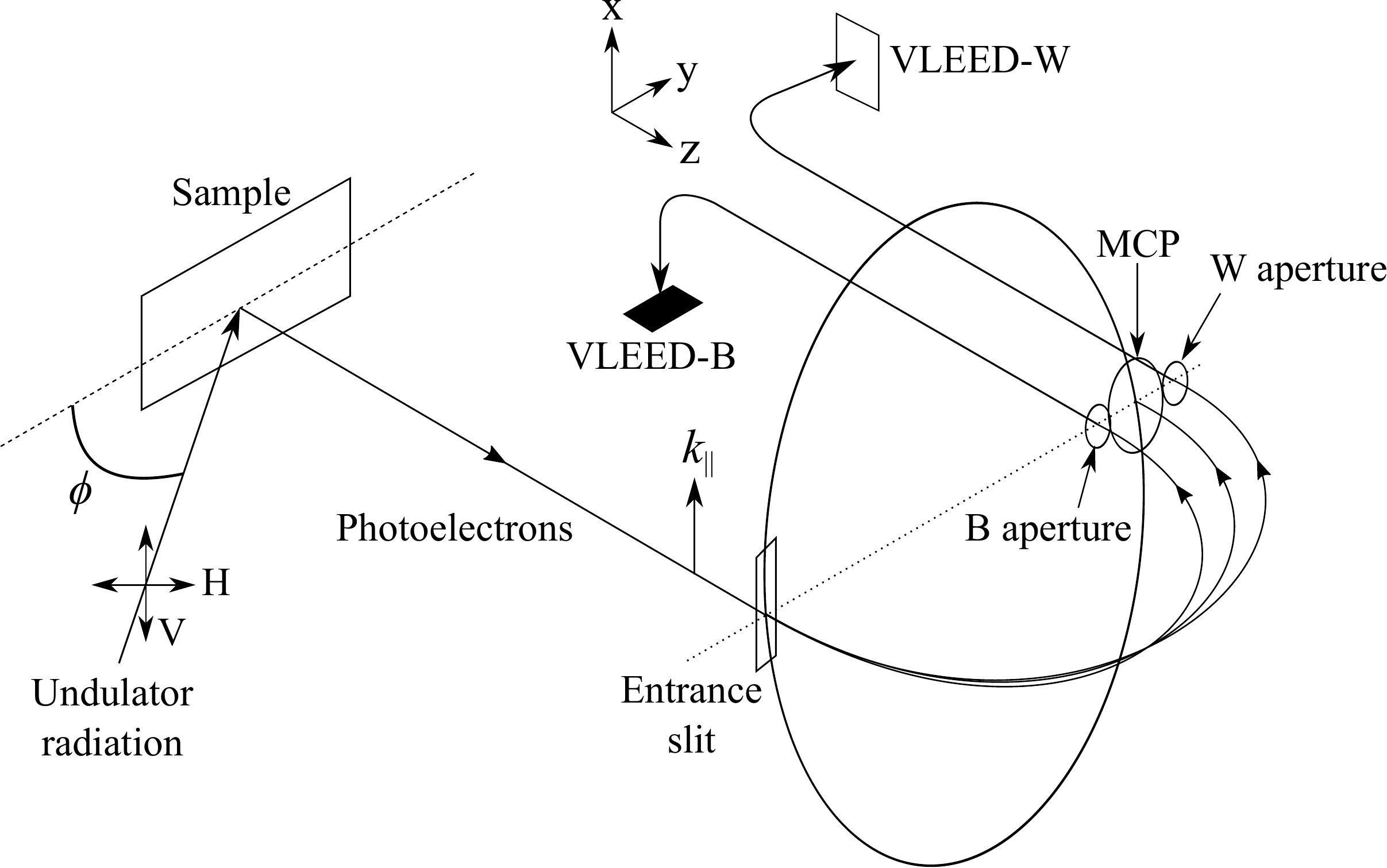}
	\caption{Schematic diagram of the end station setup at the APE-LE beamline of the Elettra Sincrotrone Trieste. Undulator radiation was either horizontally (H) or vertically (V) polarised and incident at an angle of $\phi = 45\,\si{\degree}$. The surface crystal momentum that is resolved due to the orientation of the entrance slit to the hemispherical analyser is labelled $k_{||}$. The detected spin polarisation directions are specified by the coordinate system used for the VLEED-W ($P_x, P_z$) and VLEED-B ($P_y, P_z$) polarimeters.}
	\label{Fig2}
\end{figure}

Photoemission experiments were conducted at the APE-LE beamline of Elettra Sincrotrone Trieste \cite{Panaccione,Chiara}. Figure \ref{Fig2} shows a schematic diagram of key features of the end station which was set up to conduct simultaneous spin- and angle-resolved photoemission experiments. The UHV chamber used for photoemission spectroscopy had a base pressure of $8\times10^{-11}\,\,\si{mbar}$. The undulator associated with the end station allows for the production of horizontally-, vertically- or circularly-polarised photons over the energy range of $20$ to $120\,\,\si{eV}$ \cite{Panaccione}. Both the horizontally and vertically-polarised photons are reported to have close to 100\% linear polarisation \cite{Panaccione}. In angle-resolved measurements for A-type samples, the momentum direction $[\bar{1}12]$ was resolved \textit{i.e.} along the $\bar{\Gamma}\bar{\text{A}}$ line in reciprocal space (Fig. \ref{Fig1}(b)). The corresponding resolved momentum direction for B-type samples was $[1\bar{1}2]$, \textit{i.e.} along the $\bar{\Gamma}\bar{\text{A}}'$ line in Fig. \ref{Fig1}(b). These directions were chosen because they are parallel to the step-edge directions of the corrugated terraces and have not been probed previously \cite{Sakamoto,Kim}. The $\bar{\Gamma}\bar{\text{A}}$ and $\bar{\Gamma}\bar{\text{A}}'$ lines do not correspond to symmetry axes, but in the second Brillouin zone they cross at $\bar{\text{X}}$.

Double-domain band-dispersion maps were obtained at sample temperatures $T_s$ of $77\,\,\si{K}$ and $300\,\,\si{K}$ with an experimental energy resolution $\Delta E = 55\,\,\si{meV}$. The ARPES measurements were made with horizontally-polarised photons at an energy, $\hbar\omega$, of $80\,\,\si{eV}$. Single domain surfaces were investigated to understand the effects on the surface states of changing photon energy and polarisation. These investigations were conducted at $77\,\,\si{K}$ using horizontally- and vertically-polarised photons at $60\,\,\si{eV}$ and $80\,\,\si{eV}$. A band-dispersion map of the Si(110)$1\times1$ surface was also obtained at $300\,\,\si{K}$.

Spin-resolved photoemission (SRPES) experiments were undertaken using either one of two orthogonal VLEED polarimeters, VLEED-W and VLEED-B. Both employ oxygen-passivated Fe(001)-p($1\times1$) scattering surfaces \cite{Bertacco,Okuda11,Okuda,Chiara} that are magnetised along one of two orthogonal axes referred to as the `active-scattering-axis' and identified by a subscript, for example, VLEED-W$_x$. The spin polarisation components of the photoemitted electrons, $P_x$, $P_y$ and $P_z$, are defined by the coordinate system shown next to the polarimeters in Fig. \ref{Fig2}. The spin resolving power, $S$, of the polarimeters (equivalent to the effective Sherman function in Mott polarimetry \cite{GayBook}) was taken to be 0.3 \cite{Ivana}. The energy resolutions for the spin-resolved measurements were $72\,\,\si{meV}$ and $36\,\,\si{meV}$ when pass energies of $20\,\,\si{eV}$ and $10\,\,\si{eV}$, respectively, were used. The lens mode and transfer lens aperture size produced an angular resolution of $0.75\,\si{\degree}$.

Using horizontally-polarised photons of energy $\hbar\omega = 80\,\,\si{eV}$, preliminary spin polarisation spectra were obtained with a pass energy of $20\,\,\si{eV}$ for a binding energy range of $0.0$ to $1.3\,\,\si{eV}$. All three spin components were measured. The polarisation direction of particular interest was $P_x$ as it was parallel to the surface crystal momentum. Hence $P_x$ corresponds to the longitudinal direction \cite{Modelling}. SRPES measurements performed at a pass energy of $10\,\,\si{eV}$ were obtained over the binding energy range of $1\,\,\si{eV}$ to $1.25\,\,\si{eV}$ for $P_x$. The longitudinal spin polarisation was also measured for different surface crystal momenta to determine if they exhibited a dependence on $k_{||}$. 

The sample was moved every 40 minutes during SRPES measurements to ensure that its surface was minimally affected by the photon beam. LEED and ARPES were employed to check for consistency between each surface region. LEED images were used to determine the single domain areas on the surface, and subsequent SRPES measurements were constrained to these areas to ensure only one domain was being photoexcited. Before SRPES was performed for a new surface region, band maps were obtained and the binding energies and $k_{||}$ values of all surface states were checked to match those of the previous region.

The spin-resolved data were obtained from energy-dependent intensity measurements of the number of electrons reflected by the positively ($I^+(E)$) and negatively ($I^-(E)$) magnetised iron surfaces. Polarisation values, $P(E)$, were calculated using a modified version of the polarisation equation
\begin{equation}
P = \frac{1}{S}\frac{I^+ - FI^-}{I^+ + FI^-},
\end{equation}
where $F$ is an instrumental correction factor \cite{Kessler,Yu,Elaine} and the energy-dependence notation has been omitted. Errors for the polarisations were obtained using either a weighted standard deviation or error propagation of $\sqrt{I^{\pm}}$ for greater than or fewer than 10 repeat measurements, respectively. The correction factor was either calculated point-by-point or as an energy-independent value; see supplemental information for further details \cite{SuppInfo}. Spin-resolved energy-distribution curves (EDCs) were obtained from the polarisations using the standard spin-intensity equations
\begin{equation}
I^{\uparrow} = \frac{(1+P)I}{2}, \quad I^{\downarrow} = \frac{(1-P)I}{2},
\end{equation}
where $I^{\uparrow}$ and $I^{\downarrow}$ are the spin-up and spin-down intensities and $I = I^+ + I^-$.

The correction factors were determined using
\begin{equation}
F = \frac{I^+_\text{Ta}}{I^-_\text{Ta}},
\end{equation}
where $I^+_{\text{Ta}}$ and $I^-_{\text{Ta}}$ are the intensities of an unpolarised electron beam reflected by the positively and negatively magnetised polarimeter iron surfaces, respectively. These were obtained by probing the polycrystalline Ta foil sample-retaining clips. The spin polarisation for polycrystalline Ta is expected to be zero, because the many microcrystallites are randomly oriented and their average area (diameter of $22\,\,\si{\micro m}$) is an order of magnitude smaller than the of area the beam spot ($150\times50\,\,\si{\micro m}$). Furthermore, their unpolished nature and the presence of Ta surface oxides and carbides should average any potential spin polarisation contributions to zero.

The highest precision Si longitudinal polarisation measurements were made at $77\,\,\si{K}$ with VLEED-W and were expected to be small \cite{Modelling}. Hence it was necessary to obtain particularly good statistics for the reflected intensities $I^+_{\text{Ta}}$ and $I^-_{\text{Ta}}$ generated from low temperature Ta for this polarimeter. The apparatus performance was highly optimised and the scattered intensities shown in Fig. \ref{Fig3_PolarSpin} are clearly very close to each other. These were obtained over the binding energy range $5 - 6\,\,\si{eV}$ using $\hbar\omega = 85\,\,\si{eV}$ to ensure a strong photoemission signal \cite{Penchina} and consistent photoelectron kinetic energies between Si and Ta data. The raw intensities for Ta at low temperature are shown in the upper panel of Fig. \ref{Fig3_PolarSpin}. The lower panel shows the corresponding instrumental asymmetry. The individual $I^+_{\text{Ta}}$ and $I^{-}_{\text{Ta}}$ points both have around $25,000$ counts ensuring good statistics {and the apparent residual periodic structure is a consequence of the energy step size (see the Supplemental Information for further details \cite{SuppInfo}). As a consequence of this study, it is clear that VLEED polarimeters can be optimised to measure polarisations smaller than $1\%$.
\begin{figure}[!htbp]
	\centering
	\includegraphics[width=\linewidth]{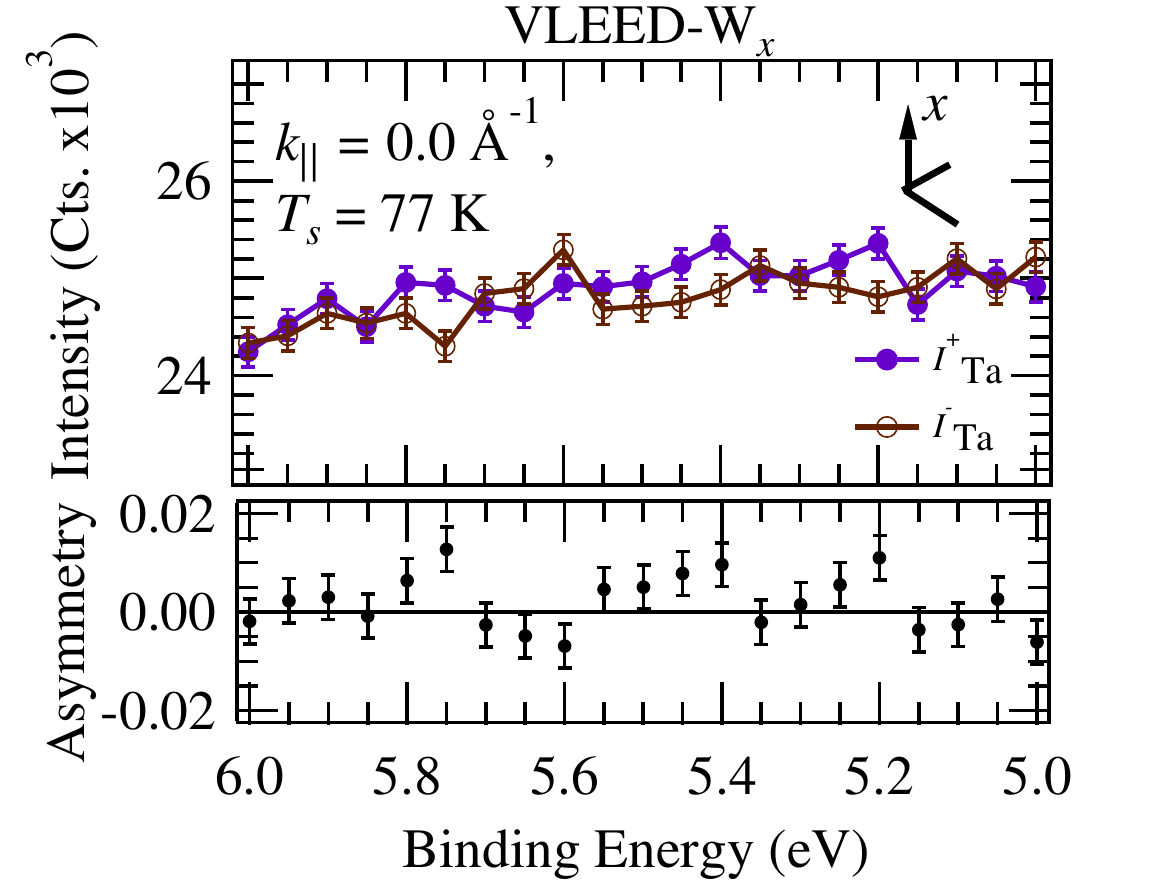}
	\caption{VLEED-W positive and negative magnetisation scattering intensities are shown by the filled (purple) and empty (brown) circles, respectively. These were obtained from a $77\,\,\si{K}$ polycrystalline tantalum surface using horizontally polarised $85\,\,\si{eV}$ photons and an energy resolution of $36\,\,\si{meV}$. The lower panel shows the corresponding instrumental asymmetry.}
	\label{Fig3_PolarSpin}
\end{figure}

\section{Results and Discussion}
\subsection{Angle-resolved photoemission} 
To investigate the validity of the AB model, band-dispersion maps of single- and double-domain surfaces were made. Figure \ref{Fig3}(a) shows a Si(110)``$16\times2$'' double domain LEED pattern for a B-type sample obtained at an electron beam energy, $E_p$, of $51\,\,\si{eV}$. LEED spots are apparent along both the $[1\bar{1}1]$ and $[\bar{1}11]$ directions showing an approximately equal mixture of L and R domains. From this sample, band maps at $77\,\,\si{K}$ (Fig. \ref{Fig3}(b)) and $300\,\,\si{K}$ (Fig. \ref{Fig3}(c)) were obtained. Energy distribution curves (EDCs) were subsequently obtained from (b) and (c) at $k_{||} = -0.5\,\,\si{\angstrom^{-1}}$ and $0.5\,\,\si{\angstrom^{-1}}$.
\begin{figure*}[!htbp]
	\includegraphics[width=0.8\linewidth]{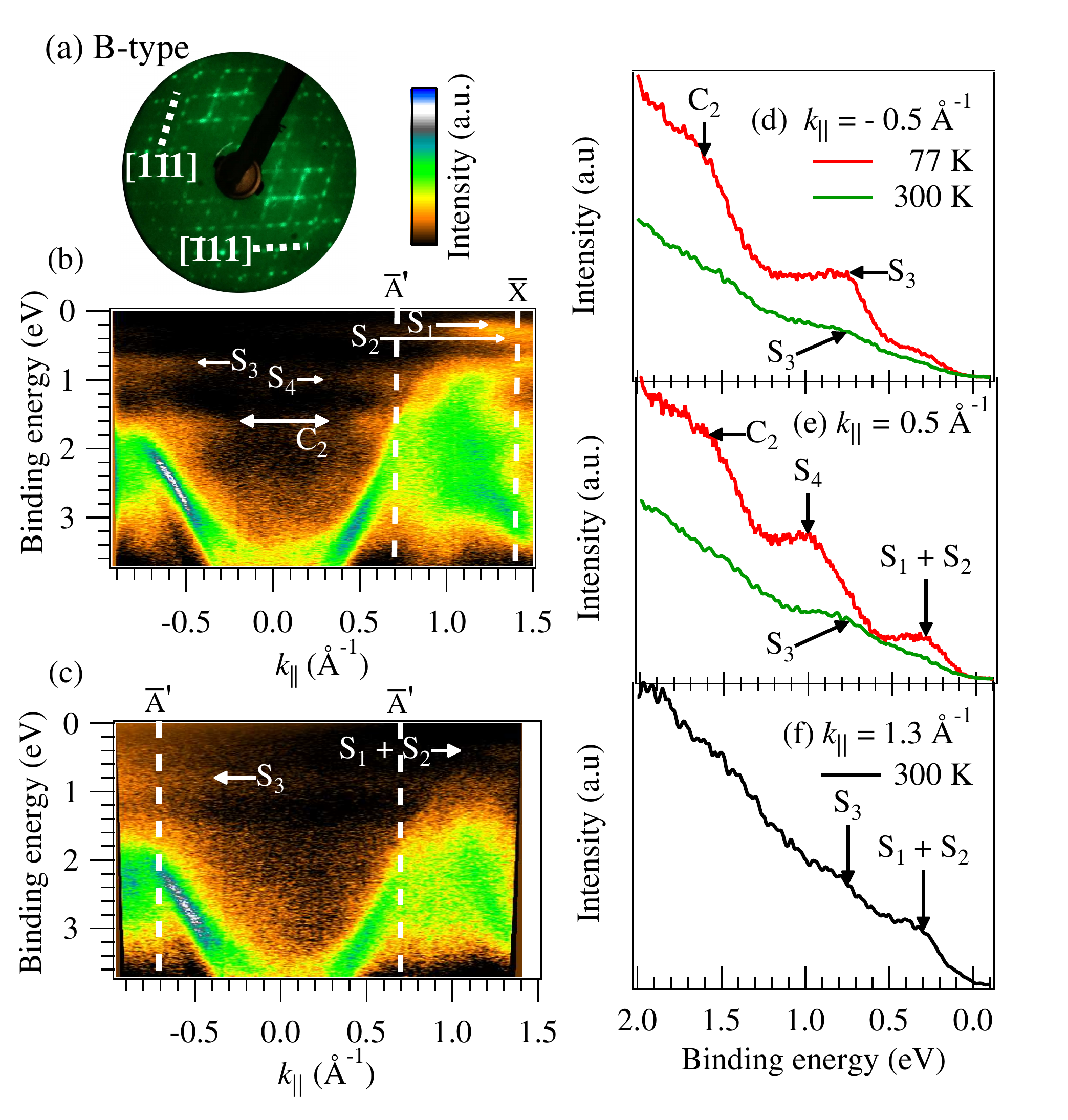}
	\caption{(a) LEED pattern showing a Si(110)``$16\times2$'' double domain for a B-type sample ($E_p = 51\,\,\si{eV}$). The crystal directions indicated are real-space lattice vectors parallel to the dashed lines. Band-dispersion maps (the intensity scale is next to the LEED pattern) were measured at $77\,\,\si{K}$, (b), and $300\,\,\si{K}$, (c), using horizontally-polarised $80\,\,\si{eV}$ photons. Energy distribution curves (EDCs) at $k_{||} = -0.5\,\,\si{\angstrom^{-1}}$ (d), and $0.5\,\,\si{\angstrom^{-1}}$ (e), were obtained from (b) and (c). The upper (red) lines in (d) and (e) correspond to measurements at $77\,\,\si{K}$ and the lower (green) lines correspond to measurements at $300\,\,\si{K}$. The EDC shown in (f) was obtained from (c) at $k_{||} = 1.3\,\,\si{\angstrom^{-1}}$.}
	\label{Fig3}
\end{figure*}

Using the same nomenclature for the surface states as Sakamoto \textit{et al.} \cite{Sakamoto}, $S_1$ to $S_4$ are identified in the band-dispersion map shown in Fig. \ref{Fig3}(b). The surface state binding energies observed are $S_1: E_B = 0.20\,\,\si{eV}$, $S_2: E_B = 0.40\,\,\si{eV}$, $S_3: E_B = 0.75\,\,\si{eV}$ and $S_4: E_B = 0.95\,\,\si{eV}$, as reported previously in Ref. \cite{Sakamoto}. Energy dispersions similar to those observed by Sakamoto \textit{et al.} \cite{Sakamoto} were also measured. The small energy separation of $0.2\,\,\si{eV}$ for the $S_1$ and $S_2$ states and their dispersions results in an intensity overlap of the photoemitted electrons causing a single surface state feature at approximately $E_B = 0.3\,\,\si{eV}$ (Fig. \ref{Fig3}(e)). The $C_2$ state, located in the bulk-band projection, is observed at $E_B = 1.6\,\,\si{\eV}$, which is similar to previous reports \cite{Sakamoto}. This is the first identification of all four surface states ($S_1$ to $S_4$) from a double-domain reconstruction.

Several of the surface states were observed at $300\,\,\si{K}$. Figure \ref{Fig3}(f) shows a peak at $E_B \approx 0.3\,\,\si{eV}$ which is due to the $S_1$ and $S_2$ states, and another peak at approximately $0.75\,\,\si{eV}$ associated with $S_3$. The $S_4$ state is not observed at $300\,\,\si{K}$, but it is clearly measured at $77\,\,\si{K}$. The temperature dependence of the $S_4$ state was observed repeatedly, but is not understood at this point. Although a shift in the binding energies of the surface states is expected between $300\,\,\si{K}$ and $77\,\,\si{K}$, all band maps show consistency in their binding energy values. Gaussian curves fitted to the $300\,\,\si{K}$ EDCs shown in Figs. \ref{Fig3}(d) and (e) are consistent with a peak at approximately $1.5\,\,\si{eV}$ which is attributed to the $C_2$ state. Low temperature EDCs, taken at approximately $40$ minute intervals and shown in the supplemental information \cite{SuppInfo}, were used to determine the longevity of the surface states in vacuo. The $S_1$, $S_2$, $S_4$ and $C_2$ states all have similar lifetimes at low temperature suggesting they are all associated with the same structure. 

Parabolic-dispersing valence band features are also evident in Figs. \ref{Fig3}(b) and (c). These occur at positive and negative values of $k_{||}$ and are connected by a minimum at $k_{||} = 0\,\,\si{\angstrom^{-1}}$. The feature at negative $k_{||}$ in both maps splits at approximately $E_B = 3\,\,\si{eV}$ and $k_{||} = -0.4\,\,\si{\angstrom^{-1}}$ into an additional minimum located at a higher binding energy (not shown on the maps). The map is not symmetric about $\bar{\Gamma}$ because the direction of the surface crystal momentum, $\bar{\Gamma}\bar{\text{A}}'$, that is being probed is not a reflection-symmetry axis of the Si(110)``$16\times2$'' surface Brillouin zones.

Figure \ref{Fig6_Rtype} shows the $77\,\,\si{K}$ band-dispersion map for an A-type single-domain Si(110)``$16\times2$'' reconstruction. The $S_1$ to $S_4$ surface states and the $C_2$ state are shown in the band-dispersion map. These were found (as expected) to be at the same binding energies as those observed in Fig. \ref{Fig3}(b), but the whole map is mirror reflected about a plane at $k_{||} = 0\,\,\si{\angstrom}^{-1}$. This is because the $\bar{\Gamma}\bar{\text{A}}'$ direction is probed in the B-type sample (Fig. \ref{Fig3}(b)) and the mirror image $\bar{\Gamma}\bar{\text{A}}$ direction is probed in the A-type sample (Fig. \ref{Fig6_Rtype}).

The same A-type single-domain Si(110)``$16\times2$'' surface was used to investigate the effects of changing light polarisation and photon energy on the band-dispersion maps; the maps shown in Figs. \ref{Fig7_80}(a) and (c) were obtained using $80\,\,\si{eV}$ horizontally and vertically-polarised photons, respectively. There are several key differences between them. Firstly, there are differences in the parabolic dispersing structure. The parabolic valence band structure that is present in Fig. \ref{Fig7_80}(c) at $k_{||} = 1.15\,\,\si{\angstrom^{-1}}$ is observed in (a) but with a reduced intensity. Thus light of orthogonal linear polarisation couples to different states in the valence band structure. Similar effects have been observed in Si(100) \cite{Uhrberg,Goldmann}.
\begin{figure}[!htbp]
	\includegraphics[width=\linewidth]{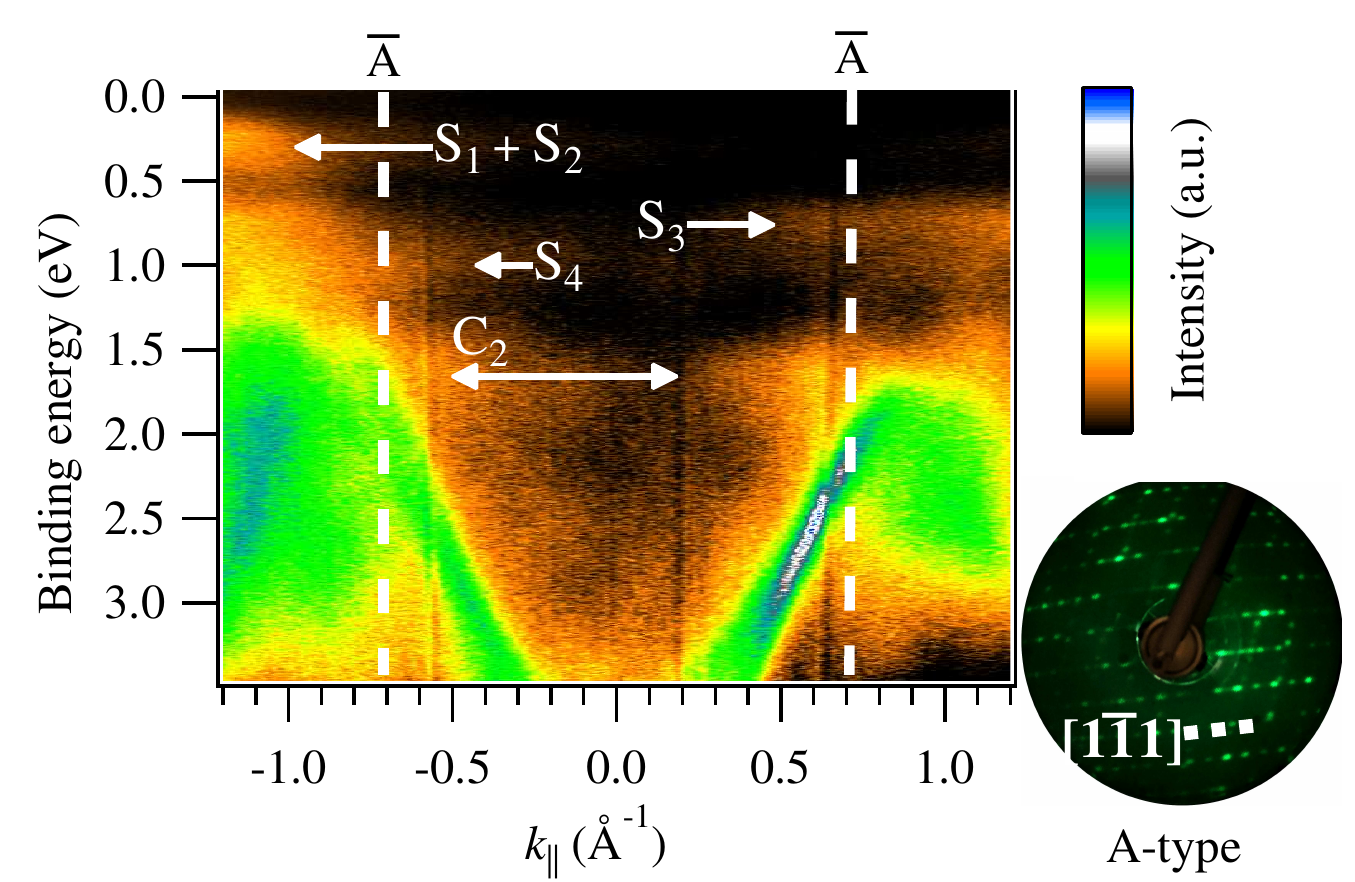}
	\caption{$77\,\,\si{K}$ band-dispersion map of an A-type single-domain Si(110)``$16\times2$'' surface. This was measured using horizontally-polarised $80\,\,\si{eV}$ photons. The lower-right inset image shows the associated single-domain LEED pattern obtained at $E_p = 50\,\,\si{eV}$. The crystal direction $[1\bar{1}1]$ is parallel to the dashed line.}
	\label{Fig6_Rtype}
\end{figure}
Secondly, the photoemission intensities of the surface states are also observed to depend on the linear polarisation of the incident light. An EDC obtained from \ref{Fig7_80}(a) is shown in Fig. \ref{Fig7_80}(b) and $S_1 + S_2$, $S_3$, $S_4$ and $C_2$ are readily identified. In contrast, an EDC derived from Fig. \ref{Fig7_80}(c) is shown in Fig. \ref{Fig7_80}(d) where the $S_1 + S_2$ and $S_3$ states are observed with a reduced intensity and $S_4$ is absent. Both band maps (Figs. \ref{Fig7_80}(a) and (c)) show an intense $C_2$ state. The downward dispersion of the $C_2$ state, as observed previously \cite{Sakamoto}, is identified using vertically-polarised light. 

\begin{figure}[!htbp]
	\includegraphics[width=\linewidth]{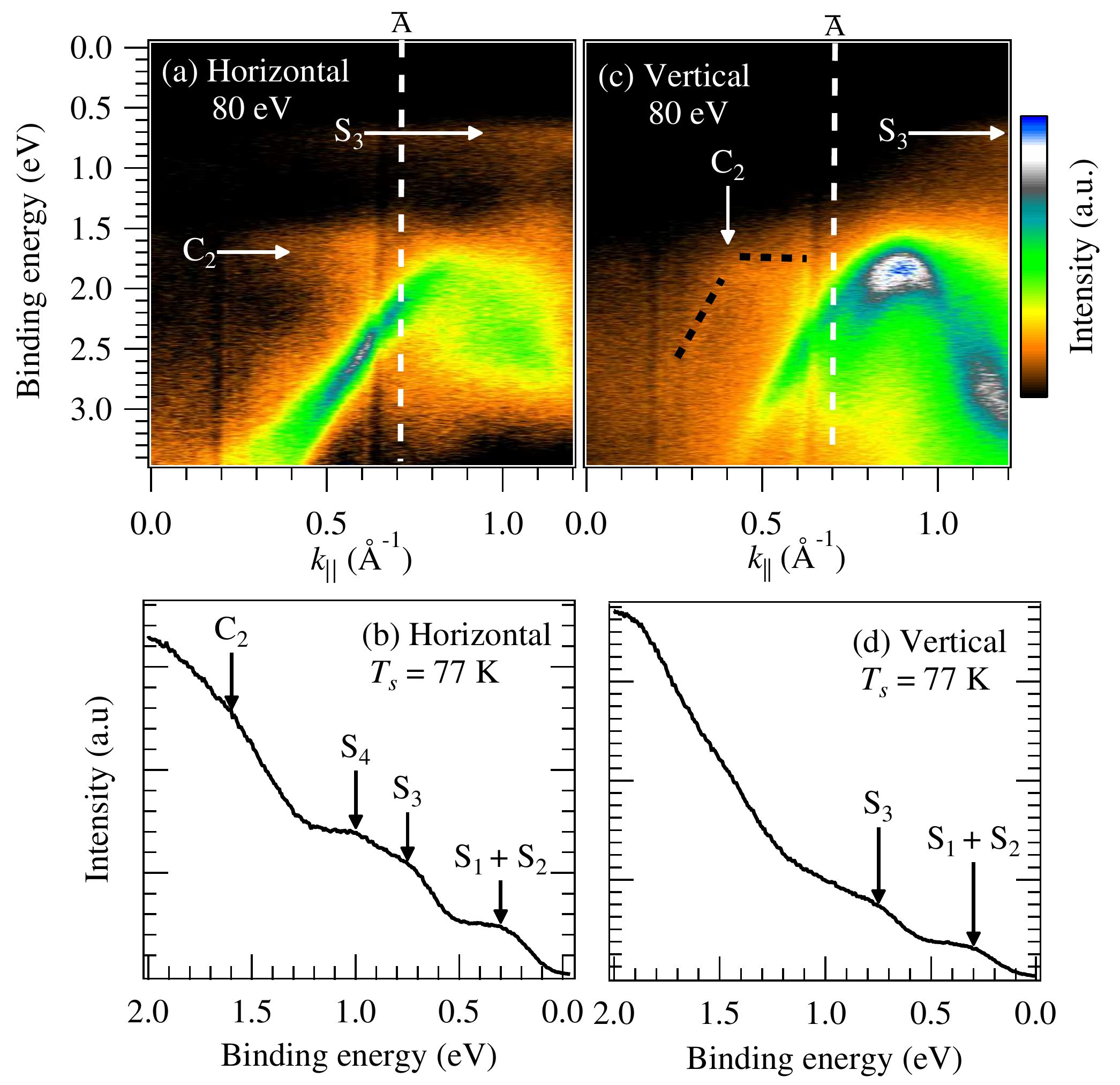}
	\caption{$77\,\,\si{K}$ band-dispersion maps of positive $k_{||}$ obtained using $80\,\,\si{eV}$ horizontally-polarised light for (a) and $80\,\,\si{eV}$ vertically-polarised light for (c). These were produced from an A-type single-domain Si(110)``$16\times2$'' surface (LEED pattern shown in Fig. \ref{Fig6_Rtype}). EDCs (b) and (d) were obtained by integrating over the surface crystal momentum axis (both $\pm k_{||}$). The black dashed line indicates the downward dispersion of $C_2$.}
	\label{Fig7_80}
\end{figure}

Band maps and EDCs were also obtained using a photon energy of $60\,\,\si{eV}$. The band-dispersion maps shown in Figs. \ref{Fig8_60}(a) and (c) were obtained from the same A-type sample (LEED pattern shown in Fig. \ref{Fig6_Rtype}) using horizontally and vertically-polarised light, respectively. Negative $k_{||}$ is presented in Fig. \ref{Fig8_60} because the intensity of $S_3$ is significantly enhanced compared to that at positive $k_{||}$ at $\hbar\omega = 60\,\,\si{eV}$. Clearly the use of vertically-polarised light attenuates the observed intensity of the $S_3$ state which is due to dipole selection rules. This is reiterated in the angle-integrated EDCs shown in Figs. \ref{Fig8_60}(b) and (d). The observed intensity of the $S_3$ state is increased using $\hbar\omega = 60\,\,\si{eV}$ compared with that observed at $\hbar\omega = 80\,\,\si{eV}$. The cause of this effect is attributed to the increased cross section of the Si $3p$ and $3s$ orbitals at $\hbar\omega = 60\,\,\si{eV}$ \cite{Yeh}. $C_2$ is more intense when horizontally-polarised light is used, but the downward dispersion of it is evident in both maps. The intensity of the other valence band features are stronger with horizontally-polarised light. For example, intensity differences are apparent in the parabolic bands.

\begin{figure}[!htbp]
	\includegraphics[width=\linewidth]{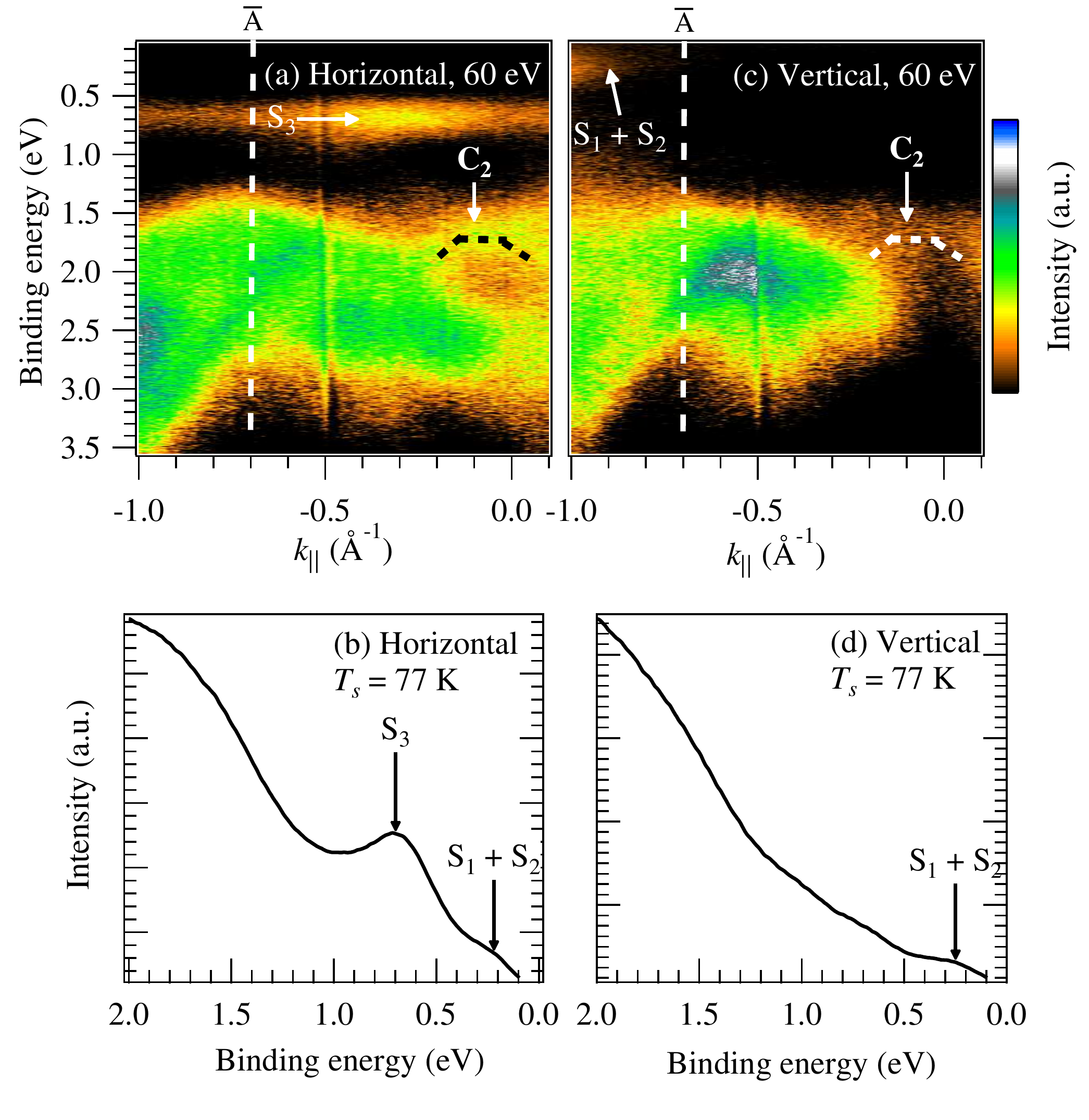}
	\caption{$77\,\,\si{K}$ band-dispersion maps of negative $k_{||}$ obtained using $60\,\,\si{eV}$ horizontally-polarised light for (a) and $60\,\,\si{eV}$ vertically-polarised light for (c). These were produced from an A-type single-domain Si(110)``$16\times2$'' surface (LEED pattern shown in Fig. \ref{Fig6_Rtype}). EDCs (b) and (d) were obtained by integrating over the surface crystal momentum axis (both $\pm k_{||}$). The black and white dashed curves indicate the downward dispersion of $C_2$.}
	\label{Fig8_60}
\end{figure}

An EDC of the Si(110)$1\times1$ surface was obtained at $300\,\,\si{K}$ and is shown in Fig. \ref{Fig4_1x1}. The upper-right inset image shows the band-dispersion map from which the EDC was derived. Only the $S_3$ state is observed in the EDC and identified at a binding energy of approximately $0.8\,\,\si{eV}$. 

\begin{figure}[!htbp]
	\includegraphics[width=\linewidth]{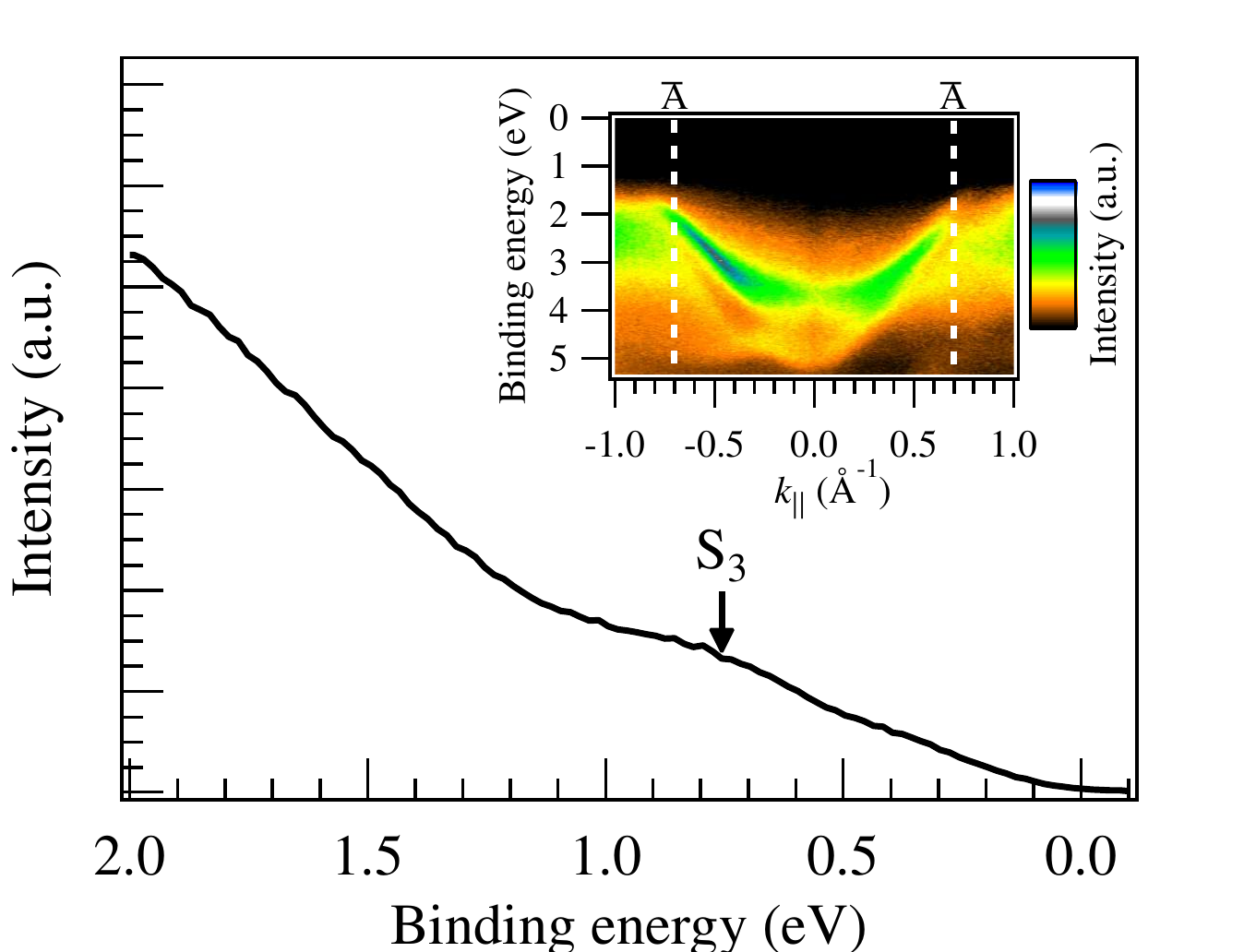}
	\caption{EDC of the Si(110)$1\times1$ surface at $300\,\,\si{K}$ using horizontally-polarised $80\,\,\si{eV}$ photons. The upper-right inset image shows the band-dispersion map from which the EDC was obtained by integrating over the $k_{||}$ axis.}
	\label{Fig4_1x1}
\end{figure}

The observations obtained from these angle-resolved photoemission experiments focusing on the surface states are summarised in Table \ref{Tab1}. All states were observed using horizontally-polarised $80\,\,\si{eV}$ photons for single and double-domain Si(110)``$16\times2$'' surfaces. The letter H is emboldened for $S_1$ - $S_4$ because the intensity of these states is significantly greater when using horizontally-polarised $80\,\,\si{eV}$ photons compared to vertically-polarised $80\,\,\si{eV}$ photons; $S_4$ was only observed using horizontally-polarised light. The surface state duration in vacuum shown in Table \ref{Tab1} (labelled as `Duration') is taken from data in the supplementary information \cite{SuppInfo}. 

The binding energy of each surface state given in Table \ref{Tab1} suggests that they can be attributed to surface dangling bonds; similar assignments have been suggested for the Si(100) surface by Goldmann \textit{et al.} \cite{Goldmann} on the basis of ARPES measurements using horizontally and vertically-polarised light. As dangling bonds are oriented mostly perpendicular to the surface, the signals from the surface states will be most pronounced using horizontally-polarised light as this has an electric field component perpendicular to the surface. This is evident from Figs. \ref{Fig7_80} and \ref{Fig8_60}. However, Table \ref{Tab1} indicates that the $S_1 + S_2$ and $S_3$ states were also visible in our spectra, but with less intensity, when using vertically-polarised light. Two reasons are proposed for this observation. First, the dangling bonds are $sp^3$ hybrids \cite{Harrison}. Thus vertically-polarised light can couple to the $s$ orbital component. The second contributing effect is that the dangling bonds could be oriented off-normal (especially for such a corrugated surface) allowing excitation using vertically-polarised light. Interestingly, $S_4$ is not observed with vertically-polarised light. This suggests the corresponding bond of $S_4$ aligns parallel to the surface normal.

Because in-plane bonds have a higher binding energy than dangling bonds, the $C_2$ state is associated with the former. Moreover, $C_2$ exhibits a shallow downward dispersion indicated by dashed lines in Figs. \ref{Fig7_80}(c), \ref{Fig8_60}(a), and \ref{Fig8_60}(b), suggesting bonding character. In addition, the intensity difference of $C_2$ between band-dispersion maps obtained with both horizontally- and vertically-polarised light is not as significant as the intensity difference for the other surface states. This suggests an orbital that is oriented mostly parallel to the surface which is formed from bulk $sp^3$ bonds. Conclusions for the states $S_1$ - $S_4$ and $C_2$ are summarised in Table \ref{Tab2}. Determination of the bonding type using the light polarisation has not been previously reported.

\begin{table}
	\begin{tabular}{ccccc}
		\hline
		\multicolumn{1}{c}{State} & \multicolumn{1}{c}{$E_B$ (eV)} & \multicolumn{1}{c}{Light} & \multicolumn{1}{c}{Surface type} & \multicolumn{1}{c}{Duration} \\
		& & polarisation & & (minutes) \\  \hline\hline
		$S_1$ & 0.20 & \textbf{H} \& V & ``$16\times2$'' & 180 \\
		$S_2$ & 0.40 & \textbf{H} \& V & ``$16\times2$'' & 180 \\
		$S_3$ & 0.75 & \textbf{H} \& V & ``$16\times2$'' \& $1\times1$ & 540 \\
		$S_4$ & 0.95 & \textbf{H} & ``$16\times2$'' & 180 \\
		$C_2$ & 1.60 & H \& V & ``$16\times2$'' & 180 \\
		\hline
	\end{tabular}
	\caption{Summary of the properties of the surface states deduced from ARPES measurements. H(V) indicates that the states were observed using horizontally(vertically) polarised light, respectively. An emboldened H indicates that the state was significantly more intense when observed with horizontally-polarised $80\,\,\si{eV}$ photons.}
	\label{Tab1}
\end{table}

Further information about the states and the structural element to which they are associated is obtained by comparing the band maps for the Si(110)``$16\times2$'' and $1\times1$ surfaces. The only surface state identified in the band map of the Si(110)$1\times1$ surface at $300\,\,\si{K}$ is $S_3$. Calculations performed by Ivanov \textit{et al.} indicate that the ideal Si(110) surface has a single state below the Fermi level \cite{Ivanov}. This suggests that the Si(110)``$16\times2$'' $S_3$ surface state should be assigned to the zigzag chains \cite{Harrison,Stekolnikov2} which are inherent to both the Si(110) planes of the ``$16\times2$'' and $1\times1$ surfaces \cite{An,Stekolnikov2}. This is further supported by the recent work of Matsushita \textit{et al.} \cite{Matsushita} which shows that $S_3$ is suppressed in the band maps of the hydrogen-terminated Si(110)$1\times1$ surface. 

\begin{table}
	\begin{tabular}{ccc}\hline 
		State & Bond type & Structural element \\ \hline\hline
		$S_1$ & Dangling bond & ``$16\times2$'' \\
		$S_2$ & Dangling bond & ``$16\times2$'' \\
		$S_3$ & Dangling bond & \,\,``$16\times2$'' \& $1\times1$: Zigzag chain \\
		$S_4$ & Dangling bond & ``$16\times2$'' \\
		$C_2$ & In-plane bond & ``$16\times2$'' \\
		\hline
	\end{tabular}
	\caption{Summary showing assignment of the structural elements and bond types of the surface states.}
	\label{Tab2}
\end{table}
We have assigned $S_1$, $S_2$ and $S_4$ to DBs and $C_2$ to an in-plane bonding state - all of which are only found on the corrugated Si(110)``$16\times2$'' reconstruction. In addition, we have assigned $S_3$ to DBs associated with the zigzag chains as this structural element is found in both the Si(110)``$16\times2$'' reconstruction and the $1\times1$ surface. These conclusions are all consistent with the AB model as within this model the $S_1$, $S_2$ and $S_4$ states are, respectively, attributed to DBs of the adatoms, the first-layer buckled-upper atoms, and the second-layer unbuckled atoms of the Si(110)``$16\times2$'' reconstructed surface. These features are only found on the Si(110)``$16\times2$'' reconstruction as a result of structural distortions resulting from the corrugated terrace structure. The AB model also associates the $S_3$ state with DBs on the unbuckled atoms of the upper zigzag chains. Some atoms at the step edges buckle upon reconstruction of the surface producing a different dangling bond state. The unbuckled atoms of the zigzag chains presented in the AB model are minimally affected by the reconstruction and are thus clearly identified with the zigzag chains in the Si(110)$1\times1$ surface. The DBs on these atoms retain the same characteristics as those found on the Si(110)$1\times1$ surface which we observed from our band-dispersion maps of the Si(110)``$16\times2$'' and $1\times1$ surfaces. Finally, the AB  model indicates that $C_2$ results from surface back bonds \cite{Sakamoto}. This is supported by our observations as $C_2$ is in the bulk-band region and observed with both horizontally- and vertically-polarised light.

\subsection{Spin-resolved photoemission}
Spin-resolved photoemission measurements were made on both A- and B-type Si(110)``$16\times2$'' surfaces that indicated good surface order (as shown by LEED) at nominal temperatures of either $300\,\,\si{K}$ or $77\,\,\si{K}$. In all cases, $80\,\,\si{eV}$ horizontally-polarised photons were used.

Initial exploratory work on an A-type single-domain Si(110)``$16\times2$'' surface at $300\,\,\si{K}$ covered the binding energy range $0.0$ to $1.22\,\,\si{eV}$ (as this encompasses the surface states $S_1$ to $S_4$). Our attention was focused on the Si surface states as these were expected to be most responsive to surface chirality effects. To maximise counts and the number of surface states probed, spin-resolved measurements were obtained at $k_{||} = 1.3\,\,\si{\angstrom^{-1}}$ where both $S_1 + S_2$ and $S_3$ are visible; a spin-integrated ARPES spectrum for this sample is shown in Fig. \ref{Fig9_SpinPrelim}(a). Spin polarisations are shown in the upper panels of Figs. \ref{Fig9_SpinPrelim}(b), (c) and (d). The lower panels in these figures show spin-resolved EDCs derived from their corresponding polarisations. Instrumental asymmetries in the various data sets were corrected for as described in the Supplemental Information \cite{SuppInfo}. A summary of the polarimeters used together with the active-scattering-axis for each of the datasets is shown in Table \ref{Table3}. The polarisations shown for Figs. 10(b), (c) and (d) are averages over the binding energies of the $S_1 + S_2$ (0.1 to 0.5 eV), $S_3$ (0.65 to 0.85 eV) and $S_4$ (0.85 to 1.05 eV) surface states. The errors shown for the polarisation values in Fig. \ref{Fig9_SpinPrelim} are only statistical; polarimeter and photon energy uncertainties have not been included \cite{Jozwiak}. 

\begin{figure}[!htbp]
	\includegraphics[width=0.8\linewidth]{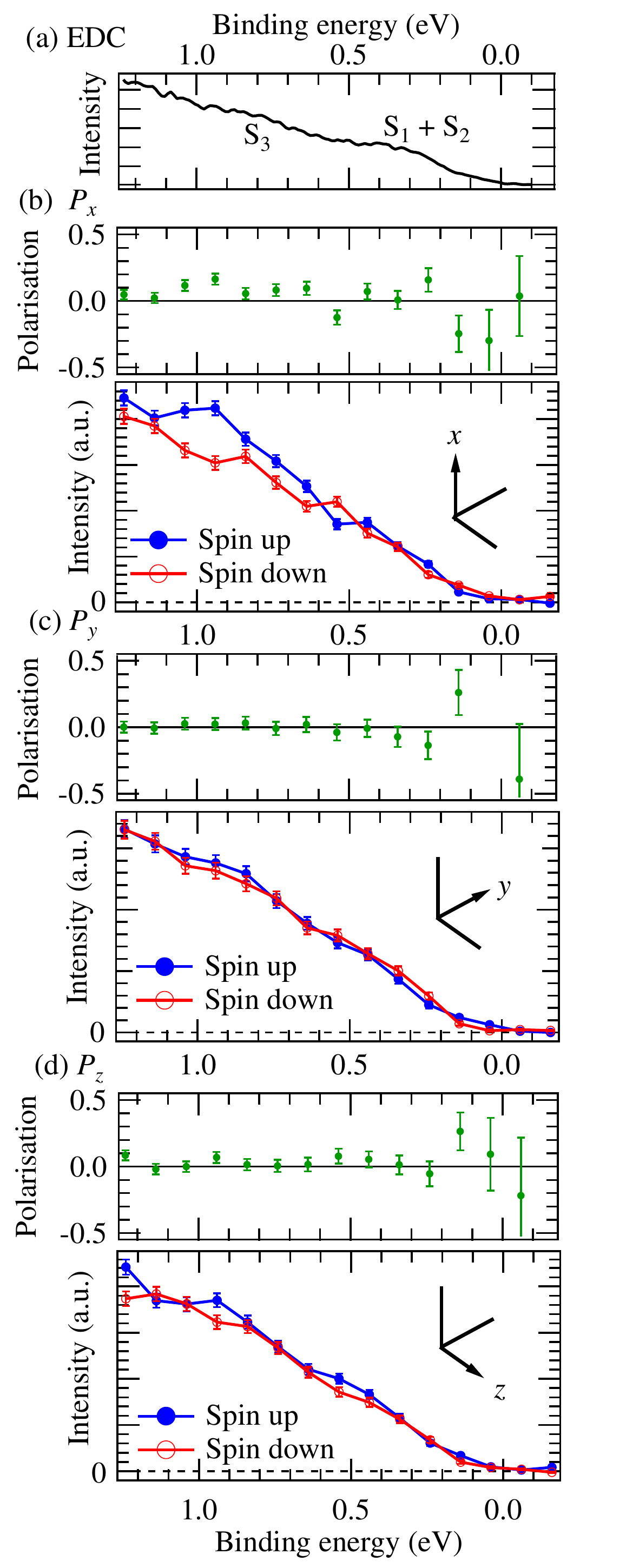}
	\caption{(a) Angle-resolved EDC for an A-type single-domain Si(110)``$16\times2$'' surface at $300\,\,\si{K}$. (b), (c) and (d) The longitudinal $P_x$, tangential $P_y$ and out of plane $P_z$ components of the spin polarisation (upper panels), and corresponding spin-resolved EDCs (lower panels). The spin-up and spin-down intensities are shown by the filled (blue) and empty (red) circles, respectively. All measurements were obtained with an energy resolution of $72\,\,\si{meV}$. Coordinate axes correspond to those shown in Fig. \ref{Fig2}.}
	\label{Fig9_SpinPrelim}
\end{figure}

\begin{table}[!htbp]
	\begin{tabular}{cccc}\hline
		Figure & Polarimeter & Active-scattering-axis & $P\,\,(\%)$ \\ \hline\hline 
		10(b) &	VLEED-W & $x$ - longitudinal & $8.9\pm1.7$ \\
		10(c) & VLEED-B & $y$ - tangential & $1.4\pm1.9$\\
		10(d) & VLEED-B & $z$ - out of plane & $1.4\pm1.8$\\
		11(a) & VLEED-W & $x$ - longitudinal & $1.9\pm0.7$\\
		11(b) & VLEED-W & $x$ - longitudinal & $-1.6\pm0.8$\\
		11(c) & VLEED-W & $x$ - longitudinal & $0.3\pm0.7$ \\ \hline
	\end{tabular}
	\caption{Polarimeter, active-scattering-axes and average polarisations for all data in Figs. \ref{Fig9_SpinPrelim} and \ref{Fig10_HighSpin}.}
	\label{Table3}
\end{table}

The tangential and out-of-plane components, $P_y$ and $P_z$, shown in Figs. \ref{Fig9_SpinPrelim}(c) and (d) respectively are statistically compatible with zero for all binding energies. The longitudinal spin component displayed in Fig. \ref{Fig9_SpinPrelim}(b) shows that, at the binding energy of $S_4$ ($0.95\,\,\si{eV}$), the polarisation values are approximately 10\%. This is anomalously large because the value is much greater than the 1-2\% obtained from our semi-empirical calculations \cite{Modelling} and the $S_4$ state is not observable at $k_{||} = 1.3\,\,\si{\angstrom^{-1}}$. One explanation for this large polarisation is that by measuring the spin at $k_{||} = 1.3\,\,\si{\angstrom^{-1}}$ we have introduced a chirality into the experimental setup. Further measurements of the longitudinal spin component were therefore made at $k_{||}$ values close to $0\,\,\si{\angstrom^{-1}}$ in order to eliminate this possibility.

\begin{figure*}[!htbp]
	\includegraphics[width=\linewidth]{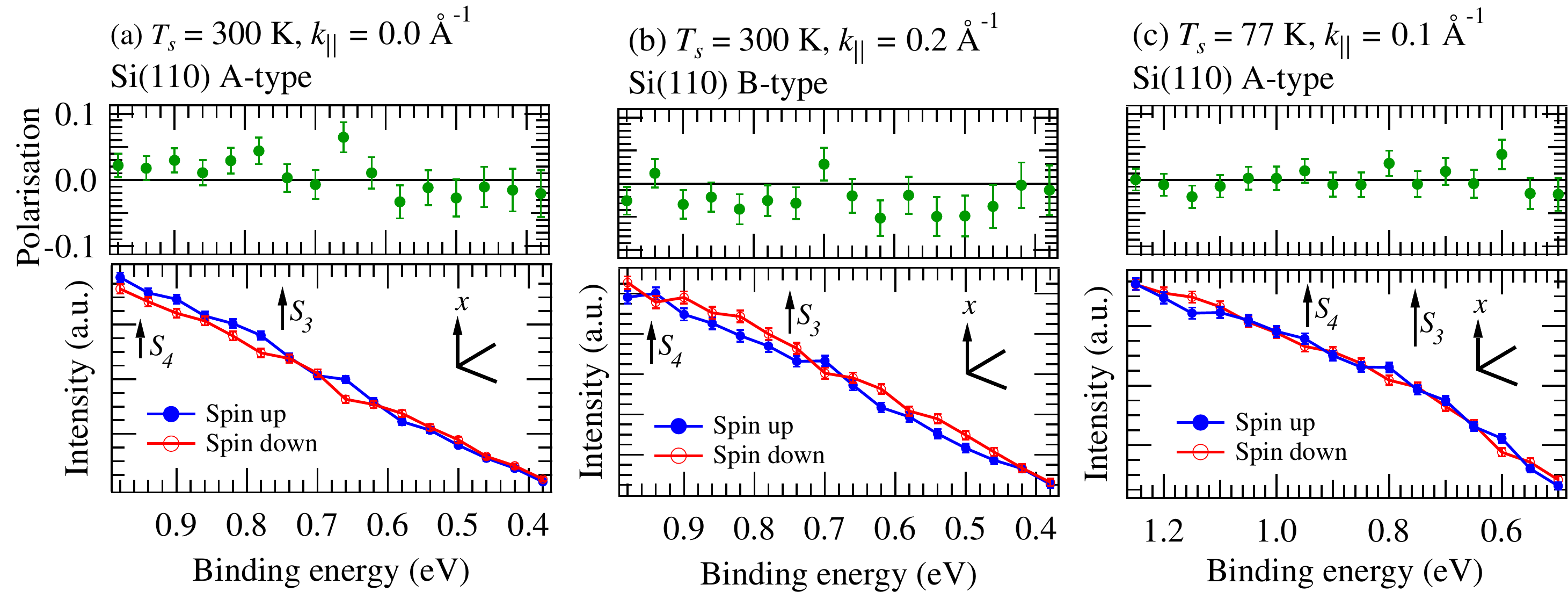}
	\caption{(a) and (b) Longitudinal spin polarisations for an A-type and B-type Si(110)``$16\times2$'' surface at $300\,\,\si{K}$ and $k_{||} = 0.0\,\,\si{\angstrom^{-1}}$ and $k_{||} = 0.2\,\,\si{\angstrom^{-1}}$ respectively. (c) Longitudinal spin polarisations for an A-type Si(110)``$16\times2$'' surface at $77\,\,\si{K}$ and $k_{||} = 0.1\,\,\si{\angstrom^{-1}}$. The corresponding spin-up (filled blue circles) and spin-down (empty red circles) intensities are shown in the lower panels. Polarisations shown in (a) and (b) were obtained with an energy resolution of $72\,\,\si{meV}$ and those in (c) obtained with a resolution of $36\,\,\si{meV}$. Coordinate axes correspond to those shown in Fig. \ref{Fig2}.}
	\label{Fig10_HighSpin}
\end{figure*}

Subsequent spin polarisation measurements displayed in Fig. \ref{Fig10_HighSpin} were focused on the $P_x$ component because it is the only one expected to invert between enantiomorphs. At $k_{||}$ close to $0\,\,\si{\angstrom^{-1}}$ only the surface states $S_3$ and $S_4$ are observed, see Fig. 6. By reducing the kinetic-energy step length and increasing the data acquisition time, more detailed $P_x$ studies were performed across the $S_3$ and $S_4$ states for A- and B-type samples. Both enantiomorphs were investigated in an attempt to observe the predicted inversion of the polarisation. Ambient temperature spin polarisations are shown in the upper panels of Figs. 11(a) and (b) for A-type and B-type Si(110)``$16\times2$'' surfaces; see Table \ref{Table3} for polarimeter collection information. The lower panels show the corresponding spin-resolved EDCs where the spin-resolved values of prime interest are at $0.75\,\,\si{eV}$ and $0.95\,\,\si{eV}$. Although not perfectly adjusted to cover the $S_4$ state, the FWHM of this state ($\sim0.2\,\,\si{eV}$) ensures that a portion of it is probed. Interestingly the average $P_x$ values for the A-type and B-type samples over $S_3$ and $S_4$, though small, have opposite signs as predicted; the A-type Si(110)``$16\times2$'' surface has $P_x = (1.9 \pm 0.7)\%$ while the B-type Si(110)``$16\times2$'' surface has $P_x = (-1.6 \pm 0.8)\%$. The removal of any unpolarised photoemission contributions from underlying bulk silicon atoms will increase these polarisation values.

In order to improve the energy resolution and reduce the randomisation of electron spins due to thermal fluctuations \cite{Bus}, the A-type sample temperature was reduced to $77\,\,\si{K}$, and the $P_x$ component was probed over the same surface states. The corresponding most accurate and precise polarisation values and derived spin-resolved EDCs are shown in the upper and lower panels of Fig. 11(c), respectively. Given the small Si $P_x$ magnitudes involved, particular effort was made to ensure optimal performance of the polarimeters and to obtain good statistics. The effective instrumental asymmetry for these measurements is shown in Figure 4; further details are given in Supplemental Information Section 2 \cite{SuppInfo}. In this low temperature case of $77\,\,\si{K}$, $P_x$ showed no discernible peaks and an average value of $(0.3\pm0.7)\%$ over the binding energy range of 0.6 to $1.1\,\,\si{eV}$. 

Our heuristic semi-relativistic calculations predict that the magnitude of the longitudinal spin polarisation for a chiral Ag lattice covered with a Bi-trimer adlayer has an average value of approximately $2.5\%$ \cite{Modelling}. Assuming that spin-orbit coupling for this alloy surface is approximately equivalent to that of Bi, then the maximum for a pure silicon surface would be expected to be less than $0.1\%$. This assumes firstly, that the longitudinal spin polarisation depends on $Z^4$ (where $Z$ is the atomic number), and secondly that the Si(110)``$16\times2$'' reconstruction has the same structure as the Bi-Ag alloy surface (which is clearly not true). 

Overall, the statistical uncertainty on the low temperature $P_x$ values over the binding energies for $S_3$ and $S_4$ leads us to give it an upper limit of 1\% consistent with the results of our calculations that indicate it should be very small. To the best of our knowledge the low temperature polarisation presented here was obtained with the lowest uncertainty yet reported for data obtained with a VLEED polarimeter, nevertheless it is clear that further reduction of the errors and improved instrumental asymmetry are still required.

\section{Conclusions}
Our angle-resolved photoemission measurements of the Si(110)``$16\times2$'' surface extend previous work by using high-resolution band mapping of double-domain, single-domain, and $1\times1$ surfaces. We assigned three of the four surface states ($S_1$, $S_2$ and $S_4$) to dangling bonds associated solely with the Si(110)``$16\times2$'' reconstruction. The remaining surface state, $S_3$, which was observed in the band maps of both the Si(110)$1\times1$ and ``$16\times2$'' surfaces, was assigned to dangling bonds on the zigzag chains of the relaxed bulk-terminated surface. The $C_2$ state observed in the bulk-band region was attributed to an in-plane bond. These assignments were produced by monitoring intensity changes of the surface states upon switching from horizontally to vertically-polarised photons. Our spectral assignments are shown to be consistent with the adatom-buckling model.

Spin-resolved photoemission measurements of the surface states for a single-domain chiral Si(110)``$16\times2$'' surface were obtained over all three polarisation components using VLEED polarimeters. First ambient temperature measurements of $P_y$ and $P_z$ gave results statistically compatible with zero polarisation but longitudinal polarisation measurements, $P_x$, for $S_3$ indicated a possible polarisation. Complementary A- and B-type samples at ambient temperature gave, as predicted, small polarisations of opposite sign in the vicinity of $S_3$ and $S_4$. However, an A-type sample was investigated further at low temperature which yielded an average polarisation of $(0.3 \pm0.7)\%$ setting an upper limit of $1\%$ for $P_x$. This value is more reliable as the polarimeter performance was highly optimised with a residual instrumental asymmetry of only $(1.9\pm1.0)\times10^{-3}$. In order to take these measurements further, higher precision polarimetry is necessary. Clearly however the chiral Si(110)``$16\times2$'' reconstruction is unlikely to be suitable for generating spin-polarised electrons in spintronic devices, although enhancement of the surface spin-orbit coupling by deposition of heavy atoms such as gold could increase the magnitude of the longitudinal spin polarisation. 

\begin{acknowledgements}
	This work was supported by EPSRC (UK) under Grant numbers EP/M507969/1 and EP/S000941/1. The research leading to these results received funding from the European Community's Seventh Framework Programme (FP7/2007-2015) under grant agreement 288879. Funding was also received from ASTeC and the Cockcroft Institute (UK), and the U.S. National Science Foundation (Awards PHY-1505794 and PHY-1430519; EB, NC and TG). The data associated with the paper are openly available from Mendeley at \url{http://dx.doi.org/10.17632/ts3284gvw5.1}
\end{acknowledgements}
\newpage
\onecolumngrid
\vspace*{1.5cm}
\begin{center}
\textbf{\LARGE Supplemental information for Spin- and angle-resolved photoemission studies of the electronic structure of Si(110)``$16\times2$'' surfaces}
\end{center} 
\vspace*{2cm}
All sections, tables, figures and equations within this supplemental information are labelled with an `S' prefix to distinguish them from those in the accompanying paper.
\section{Surface state duration in vacuum}
The durations of states $S_1$ - $S_4$ and $C_2$ were determined in ultra-high vacuum at $77\,\,\si{K}$ using angle-resolved photoemission. Here `duration in vacuum' corresponds to the time between the sample reconstruction and the measurement of angle-resolved data that shows the intensity of a state falling below the background electron counts. The measured surface-state photoemission intensities are adversely affected by both adsorption of residual gas atoms and by the photon beam.  In order to gain some insight into the relative importance of these factors the sample was transferred from the deposition chamber, in which the reconstruction was performed, to the photoemission chamber. It was then cooled to $77\,\,\si{K}$ as quickly as possible, which took about 30 minutes. Once aligned correctly in the photoemission chamber an initial band map was obtained from which an EDC was derived by integrating over all $k_{||}$ and the status of states $S_1$ - $S_4$ and $C_2$ determined. Subsequent band maps and EDCs were recorded at approximately 40 minute intervals and were obtained using the same method; after each interval the sample was translated by a small amount. By doing this the effect of the residual gases increased continuously but the effect of the photon beam was regularly reset. Therefore, the state duration values are primarily affected by surface contaminations.

\begin{figure}[!hptb]
	\centering
	\includegraphics[width=0.6\linewidth]{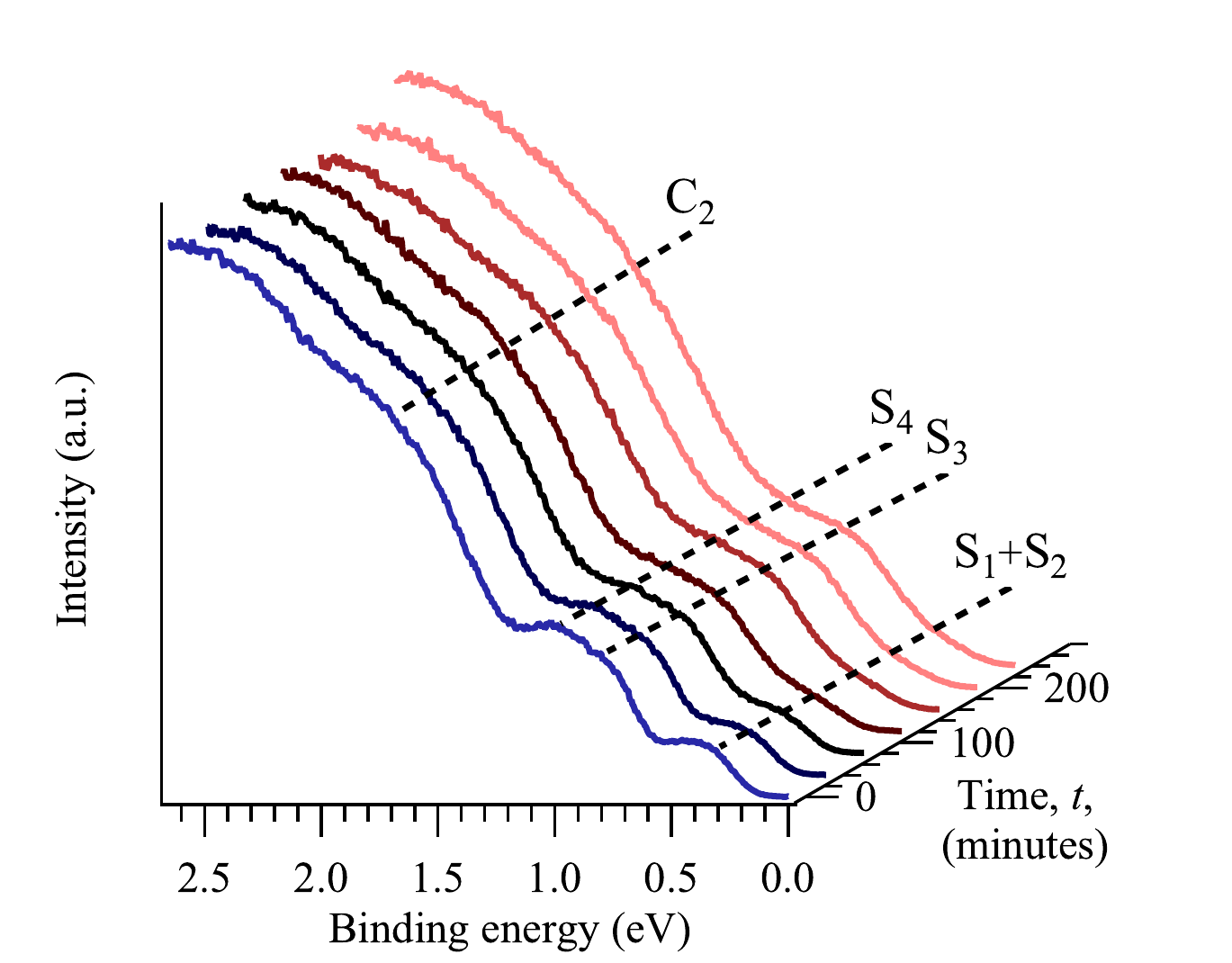}
	\caption{EDCs showing the time evolution of the surface states of the Si(110)``$16\times2$'' surface at $77\,\,\si{K}$. The isoenergetic dashed lines are used to indicate the maximum intensity of the surface states over time.}
	\label{Fig1_RT}
\end{figure}

The resultant EDCs are shown in Fig. \ref{Fig1_RT}. The EDC at $t$ = 0 minutes (i.e. the time of the first angle-resolved photoemission measurement after surface reconstruction) was produced from the band map shown in Fig. 5(b) of the accompanying paper. An extra 30 minutes has been added to the state durations below to account for the time required to move the sample from the preparation stage to the photoemission stage. The durations of the $S_1+S_2$, $S_4$ and $C_2$ states measured at $T_s = 77\,\,\si{K}$ were found to be approximately 180 minutes. The duration of the $S_3$ state was found to be three times longer than the other states at 540 minutes. The longer duration of $S_3$ is attributed to its presence in both the $1\times1$ and ``$16\times2$'' surfaces.

\section{Si data collection and analysis}
Reported here are details of the acquisition process and correction procedure for the six Si data sets reported in Figs. 10 and 11. These data were obtained in 40 minute time slots that involved an initial ARPES scan to check the surface state intensities. Following this, SRPES measurements were made in which the polarimeter scattering surface magnetisation along the active-scattering-axis was switched six times in the order +-/-+/+-. After the 6 measurements, the sample was repositioned within the single-domain region. The ARPES check and SRPES measurements were then repeated. Each set of measurements were obtained with a single polarimeter, for example VLEED-W, along an active-scattering axis, such as $x$, both represented as VLEED-W$_x$.  

Table \ref{STab1} lists the type of corrections and other key parameters relevant to each of the Si spin-resolved data sets. The corrections factors shown are averages over the whole binding energy range of each data set. The polarisations shown for Figs. 10(b), (c) and (d) are averages over the binding energy ranges of $S_1 + S_2$, $S_3$ and $S_4$, those shown for Figs. 11(a), (b) and (c) are only averages over $0.65$ to $1.05\,\,\si{eV}$ encompassing the $S_3$ and $S_4$ states. 

\subsection{Correction of Si polarisations}

Spin polarisations for all of the Si measurements reported in the accompanying paper were corrected by removing instrumental asymmetries that were apparent in their spin-up and spin-down intensities. Obtained with a step size of $100\,\,\si{meV}$, Figs. \ref{SFig2}(a), (b) and (c) show respectively the raw Si data (for Fig. 10(c)), the Ta data set used to correct the instrumental asymmetries and the corrected Si data. With the smaller step size of $40\,\,\si{meV}$, Figs. \ref{SFig2}(d), (e) and (f) show respectively the raw Si data (for Fig. 11(b)), the Ta data used to correct them and the Si results after correction. Only two types of instrumental asymmetry were observed and point-by-point correction factors, $F(E)$, where $E$ is the electron binding energy, were employed to reduce the magnitude of both effects. The exception to this was the Si data set shown in Fig. 11(c) where an energy-independent correction factor was used to remove the instrumental asymmetry.

The correction factor was determined using unpolarised photoelectrons produced from the thin strips of Ta foil used to retain the sample. All of the experimental parameters (apart from the sample lateral position and photoelectron beam focus) were kept constant for both the Si and Ta data. 

\begin{turnpage}
	\begin{table}
		\centering
		\begin{tabular}{c|c|c|c|c|c|c|c}
			Data Set & Polarimeter & Active-scattering-axis & Correction type & Average Correction & Polarisation, $P$ (\%) & $T$ (K) & Step size \\ 
			Figure- & VLEED- & & &Factor, $F$  &  & & (meV)\\ \hline \hline
			& & & & & & & \\
			10(b) & W & $x$ (longitudinal) & point-by-point & $1.042\pm0.004$ & $8.9\pm1.7$ & 300 & 100 \\
			& & & & & & & \\
			10(c) & B & $y$ (tangential) & point-by-point & $0.929\pm0.004$& $1.4\pm1.9$ & 300 & 100\\
			& & & & & & & \\
			10(d) & B & $z$ (out-of-plane) & point-by-point & $0.989\pm0.004$ & $1.4\pm1.8$ & 300 & 100 \\
			& & & & & & & \\
			& & & & & & & \\
			11(a) & W & $x$ (longitudinal) & point-by-point & $1.050\pm0.008$ & $1.9 \pm 0.7$ & 300 & 40 \\
			& & & & & & & \\
			11(b) & W & $x$ (longitudinal) & point-by-point & $1.070\pm0.008$ & $-1.6 \pm 0.8$ & 300 & 40 \\
			& & & & & & & \\
			11(c) & W & $x$ (longitudinal) & energy independent &$1.004\pm0.002$ & $0.3\pm0.7$ & 77 & 40 \\
		\end{tabular}
		\caption{Key parameters and correction information for each data set reported in the accompanying paper, where $T$ corresponds to the sample temperature during ARPES and SRPES measurements.}
		\label{STab1}
	\end{table}
\end{turnpage}

\begin{turnpage}
	\begin{figure}[!htbp]
		\centering
		\includegraphics[width=\columnwidth]{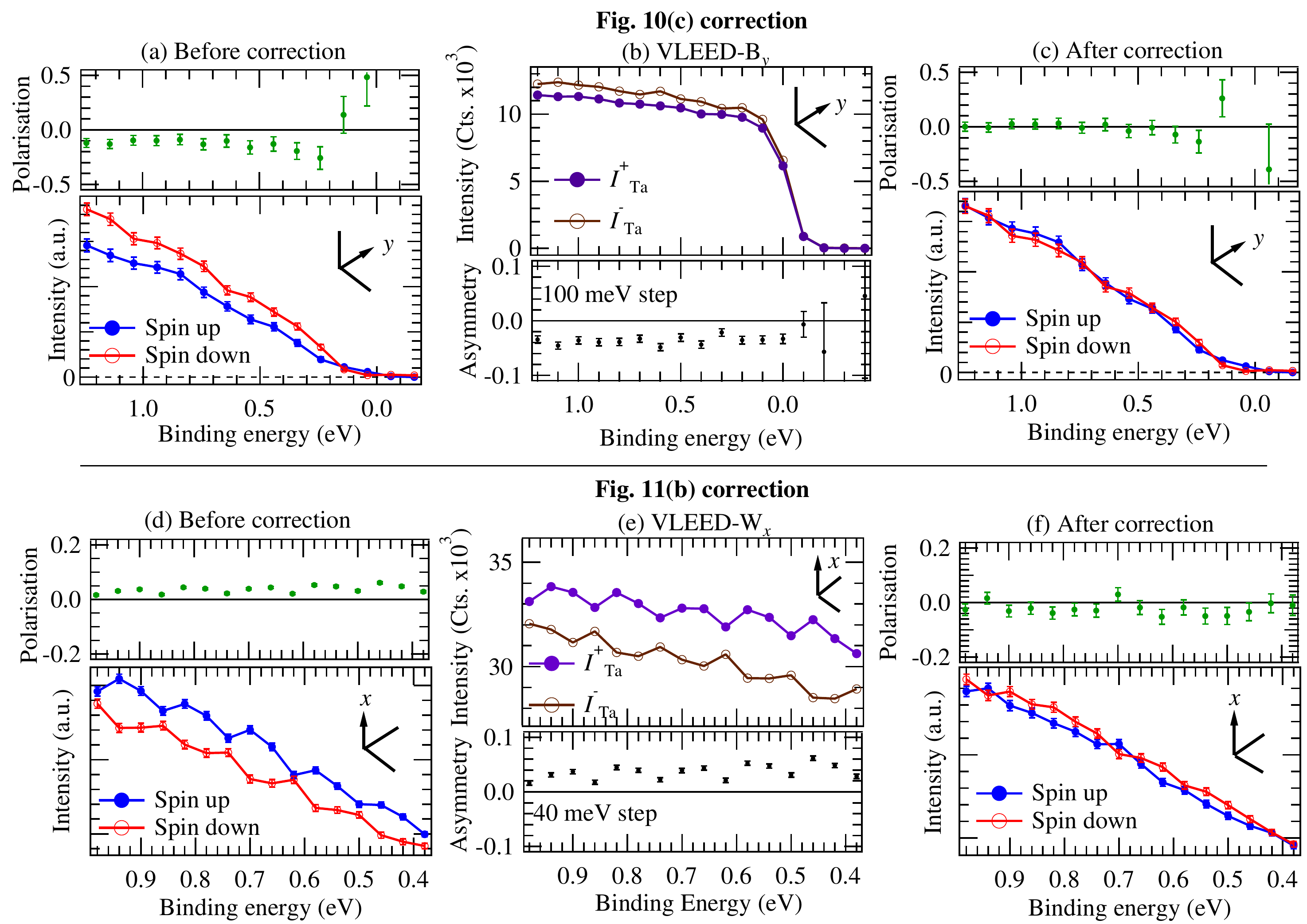}
		\caption{Upper and lower halves show the correction steps for the data sets shown in Figs. 10(c) and 11(b), respectively. (a), (c), (d) and (f): Spin polarisations and spin-resolved EDCs obtained at $300\,\,\si{K}$ from type-A and type-B Si samples, where (a) and (d) show the data sets ``Before correction'' and (c) and (f) show the data sets ``After correction''. Polarisations are shown in the upper panels; spin-up and spin-down intensities are shown in the lower panels by the filled (blue) and empty (red) circles, respectively. (b) and (e): Raw polarimeter scattering intensities and corresponding asymmetries for polycrystalline Ta recorded at $300\,\,\si{K}$. The coordinate system corresponds to that shown in Fig. 3.}
		\label{SFig2}
	\end{figure}
\end{turnpage}

Well prepared, atomically flat, single crystal surfaces of Ta can exhibit sizeable spin polarisations (see in particular \cite{Henk,Wortelen}). However, our spin detector correction measurements were performed using a Ta foil - which was not highly polished and which was polycrystalline with an average grain size of around $22\,\,\si{\micro m}$ (smaller than the photon spot size). In addition, it undoubtedly had a random crystallite orientation and had significant amounts of carbon and oxygen on it. All of these factors combine to give rise to a zero spin polarisation signal and any measured deviation from zero asymmetry was attributed to instrumentally derived effects.

To determine the polarimeter's instrumental asymmetry for a given binding energy and hence the correction factor $F(E)$, the intensity of the scattered Ta derived photoelectron beam was measured 6 times using a magnetisation order reversal of +-/-+/+- to avoid time dependent effects; this measurement procedure was repeated twice. Repeat scattered intensities from the Fe surface in the polarimeters were summed to give $I^+_{\text{Ta}}(E)$ and $I^-_{\text{Ta}}(E)$. The instrumental asymmetries, $A_i$, were then calculated from these intensities using 
\begin{equation}
A_i = \frac{I^+_{\text{Ta}}(E) - I^-_{\text{Ta}}(E)}{I^+_{\text{Ta}}(E) + I^-_{\text{Ta}}(E)}.
\end{equation}
The corresponding correction factors were calculated using
\begin{equation}
F(E) = \frac{I_\text{Ta}^+(E)}{I_\text{Ta}^-(E)}.
\end{equation}

Typical instrumental asymmetry measurements where the step size corresponds to $100\,\,\si{meV}$ are shown in Fig. \ref{SFig2}(b), those where the step size is $40\,\,\si{meV}$ are shown in Fig. \ref{SFig2}(e). In both figures, the summed scattered intensities and corresponding instrumental asymmetries are shown in the upper and lower panels, respectively. On average, the counts for both $I^+_{\text{Ta}}(E)$ and $I^-_{\text{Ta}}(E)$ were approximately 12,000 for the data shown in Fig. \ref{SFig2}(b) and approximately 30,000 for the data shown in Fig. \ref{SFig2}(e).

The two instrumental effects are clear in Figs. \ref{SFig2}(b) and (e). The first is observed in Fig. \ref{SFig2}(b) where the asymmetry values are shifted from zero, they have the same sign and they reduce slightly towards a binding energy of zero. The second is observed in the intensities and asymmetries shown in Fig. \ref{SFig2}(e) obtained with a step size of $40\,\,\si{meV}$. Not only are the asymmetry values all shifted with the same sign away from zero, but the scattered intensities show there is a repeating antiphase periodic structure with a period of approximately $120\,\,\si{meV}$. Both of the instrumental effects arise due to systematic errors caused by misalignment of the incident electron beam on the Fe(001) scattering surfaces in the polarimeters. The offset in scattered intensities are the result of stray magnetic field changes that occur on polarimeter target magnetisation reversal and which are slightly asymmetrical with respect to the scattering position. The periodic oscillations result because deflectors are used in the lens system of the electron analyser, the voltages of which are controlled by a lens table which ensures optimum beam alignment approximately every $120\,\,\si{meV}$ \cite{Ivana}. With a step size of only $40\,\,\si{meV}$, the beam position is only re-optimised after every third step. Therefore, after the first and second steps the beam position is slightly sub-optimal with respect to beam position and stray fields, resulting in the observed scattering intensity fluctuations. With a photoelectron step size of $100\,\,\si{meV}$ the system minimises very effectively any changes in the incident beam position.

Polarisation values, $P$, for Si data obtained were calculated using
\begin{equation}
P = \frac{1}{S}\frac{I^+(E) - F(E)I^-(E)}{I^+(E) + F(E)I^-(E)},
\end{equation}
where $I^+(E)$ and $I^-(E)$ correspond to the Si-derived photoelectron beam intensities reflected by the polarimeter's positively and negatively magnetised Fe target, respectively, and $F(E)$ is the energy-dependent correction factor. In all cases, a spin sensitivity factor, $S$, of $0.3$ was used.

Two approaches were used to obtain Si spin-resolved photoemission measurements. In the first approach, an acquisition time of $1.5\,\,\si{s}$ per point, a binding energy range of $1.4\,\,\si{eV}$ and a step size of $100\,\,\si{meV}$ were used. In the second approach, an increased data acquisition time of $2.5\,\,\si{s}$ per point, a reduced binding energy range of $0.6\,\,\si{eV}$ and a decreased step size of $40\,\,\si{meV}$ were used. The key difference between these approaches was the binding energy step size. The two instrumental effects that are evident in the Si data of Figs. \ref{SFig2}(a) and (d), the negative bias to the polarisation data and the periodic structure, have been successfully reduced by using the Ta data sets in Figs. \ref{SFig2}(b) and (e), respectively.

For the high-precision low temperature data reported in Fig. 11(c), a single energy-independent correction factor very close to 1 (see Table \ref{STab1}) was used to correct the raw data. This $F$ factor was calculated by averaging the polarimeter scattering intensities, $I^+_{\text{Ta}}(E)$ and $I^-_{\text{Ta}}(E)$ shown in Fig. 4, that were obtained with $85\,\,\si{eV}$ photons and electrons with binding energies between $5$ and $6\,\,\si{eV}$. This binding energy range was chosen in order to achieve a good signal-to-noise ratio for the correction factor but keep the Ta photoelectron kinetic energies close to those of the Si. The asymmetries are distributed about zero but still show a small residual periodic structure with a magnitude of approximately 0.015 as a consequence of using a $40\,\,\si{meV}$ step size.

\bibliographystyle{naturemag}

\end{document}